\documentclass[conference]{IEEEtran}
\IEEEoverridecommandlockouts
\usepackage{float}
\usepackage{cite}
\usepackage{amsmath,amssymb,amsfonts}
\usepackage{algorithm}
\usepackage{algpseudocode}
\usepackage{graphicx}
\usepackage{float} 
\usepackage{textcomp}
\usepackage{xcolor}
\usepackage{graphics}
\usepackage{epstopdf}
\usepackage{caption}
\usepackage{tabularx}
\usepackage{amssymb}
\usepackage{diagbox}
\def\BibTeX{{\rm B\kern-.05em{\sc i\kern-.025em b}\kern-.08em
    T\kern-.1667em\lower.7ex\hbox{E}\kern-.125emX}}
\allowdisplaybreaks[3]
\makeatletter
\renewcommand{\maketag@@@}[1]{\hbox{\m@th\normalsize\normalfont#1}}%
\makeatother

\begin{document}

\title{Secure Satellite Communications via Multiple Aerial RISs: Joint Optimization of Reflection, Association, and Deployment}

\author{
Zhaole Wang,
Naijin Liu,
Xiao Tang,~\IEEEmembership{Member,~IEEE,}
Shuai Yuan,
\\Chenxi Wang,
Zhi Zhai,~\IEEEmembership{Member,~IEEE,}
Qinghe Du,~\IEEEmembership{Member,~IEEE,}
and Jinxin Liu,~\IEEEmembership{Member,~IEEE}
\thanks{Z. Wang is with the School of Future Technology, Xi'an Jiaotong University, Xi'an 710049, China. (e-mail: wangzhaole@stu.xjtu.edu.cn).}
\thanks{N. Liu is with the Institute of Telecommunication and Navigation Satellites, China Academy of Space Technology, Beijing 100094, China. (e-mail: liunaijin@xjtu.edu.cn).}
\thanks{Xiao Tang is with the School of Information and Communication Engineering, Xi'an Jiaotong University, Xi'an 710049, China, and also with Shenzhen Research Institute of Northwestern Polytechnical University, Shenzhen 518057, China. (e-mail: tangxiao@xjtu.edu.cn)}
\thanks{S. Yuan is with the Qian Xuesen Laboratory of Space Technology, China Academy of Space Technology, Beijing 100094, China. (e-mail: yuanshuai\_casc@163.com).}
\thanks{Q. Du is with the School of Information and Communication Engineering, Xi'an Jiaotong University, Xi'an 710049, China. (e-mail: duqinghe@mail.xjtu.edu.cn).}
\thanks{C. Wang, Z. Zhai and J. Liu are with the School of Mechanical Engineering, Xi’an Jiaotong University, Xi’an 710049, China. (e-mail: wangchenxi@xjtu.edu.cn, zhaizhi@xjtu.edu.cn, jinxin.liu@xjtu.edu.cn).}
}

\maketitle

\begin{abstract}
Satellite communication is envisioned as a key enabler of future 6G networks, yet its wide coverage with high link attenuation poses significant challenges for physical layer security. In this paper, we investigate secure multi-beam, multi-group satellite communications assisted by aerial reconfigurable intelligent surfaces (ARISs). To maximize the sum of achievable multicast rates among the groups while constraining wiretap rates, we formulate a joint optimization problem involving transmission and reflection beamforming, ARIS-group association, and ARIS deployment. Due to the mixed-integral and non-convex nature of the formulated problem, we propose to decompose the problem and employ the block coordinate descent framework that iteratively solves the subproblems. Simulation results demonstrate that the proposed ARIS-assisted multi-beam satellite system provides a notable improvement in secure communication performance under various network scenarios, offering useful insights into the deployment and optimization of intelligent surfaces in future secure satellite networks.
\end{abstract}


\section{Introduction}
The explosive growth of connected devices with diverse form factors worldwide, coupled with surging demands for high data rates and seamless service continuity, has driven the development of non-terrestrial networks (NTN) in fifth-generation (5G) and emerging sixth-generation (6G) wireless systems. Satellite communication systems are critical to enable ubiquitous connectivity and enhance network capacity in 6G ecosystems, owing to their inherent advantages in scalability, reliability, and global coverage~\cite{a1,a2,a3}. To address the escalating requirements for broadband interactive services, multibeam satellites have emerged as a viable solution for delivering spectrum-efficient and high-throughput data transmissions~\cite{a4}. 

Although satellite communication plays a significant role in 6G communication, its broader coverage compared to terrestrial communication exposes it to more severe information security risks, making it more vulnerable to interception threats. As a promising solution to this challenge, physical layer security has gained significant attention due to its keyless operation and potential for perfect secrecy~\cite{a5,a6}. Although physical layer security has achieved highly effective security assurances in wireless communication systems, the physical layer security of satellite communication still faces significant challenges due to the following reasons. The channel advantage of the legitimate side, which is essential for ensuring physical layer security, can hardly be maintained, as the channel difference between legitimate users and eavesdroppers diminishes significantly. Additionally, low-Earth orbit (LEO) satellites experience weak signal strength caused by long-distance propagation. In such scenarios, the deterioration of physical layer security performance may become inevitable~\cite{a7}.

Meanwhile, the reconfigurable intelligent surfaces (RIS) has emerged and gained rapid attention in the field of wireless communications~\cite{a8}. RIS typically comprises numerous passive electromagnetic elements that reflect signals and adjust wireless channels through precisely designed phase shifts of each element~\cite{a9}. Currently, RIS has been employed to enhance wireless communication across various dimensions, including system throughput, energy efficiency, network coverage~\cite{R1}, and beyond~\cite{a10,a11}. Given its role in influencing wireless propagation, RIS has emerged as a promising solution for significantly enhancing physical layer security performance. By dynamically optimizing the phase and amplitude of electromagnetic waves through programmable passive reflection units, RIS actively shapes wireless channels to amplify legitimate signals and suppress eavesdropper reception. Moreover, RIS has also led to substantial advancements such as secure transmission beamforming, artificial noise design, anti-jamming communications, and secret key generation~\mbox{\cite{a12,a13,a14,a15}}. 

The potential of RIS provides an effective approach to enhance physical layer security in satellite communication systems~\cite{a19}. While the system security performance can be largely affected by optimizing reflection beamforming, the quality of the reflection link is determined by the induced channel and the reflection channel. Thus, the reflection location is crucial for determining overall performance~\cite{a20}. In this regard, researchers have suggested that RIS can be easily mounted onto aerial platforms such as unmanned aerial vehicles, balloons, or high-altitude platforms, leading to the concept of aerial RIS (ARIS)~\cite{a21,a22}. Compared to the conventional terrestrial RIS, ARIS offers flexible mobility to further improve the secure transmission performance. By reflecting signals from optimal positions, ARIS can significantly reduce signal leakage and enhance the secrecy rate~\cite{a23}. Although prior work has explored the use of RIS in satellite communications, these studies have primarily focused on terrestrial RIS and single ARIS deployments. However, due to the large coverage area of satellite beams and the dual fading effect in the reflection channel, a single ARIS may not suffice to meet the extensive coverage needs of satellite communication. In such scenarios, deploying multiple ARISs across each beam emerges as a viable solution to ensure comprehensive user coverage while mitigating signal degradation.

Given the advantages of ARIS with flexible deployment and reflection-enhanced communications, we in this paper aim to employ multiple ARISs to enhance the security performance of multi-group multi-beam satellite communications. The main contributions of this paper are summarized as follows:
\begin{itemize}
\item We propose an ARIS-assisted secure satellite communication framework that enhances physical layer security by jointly exploiting the reflection, association, and deployment of multiple ARISs. We formulate a joint optimization problem to maximize the sum multi-cast rate across all groups while constraining the wiretap rates of multiple eavesdroppers.
\item To address the computational complexity of the formulated mixed integral and non-convex problem, we develop a block coordinate descent (BCD) framework that decomposes the optimization into tractable subproblems. The transmission and reflection beamforming is obtained through semidefinite programming (SDP), and the association and deployment subproblems are solved with successive convex approximation (SCA), with penalty mechanism introduced to tackle the integral constraints.
\item The simulation results across a variety of scenarios are provided to corroborate the effectiveness of our proposal, highlighting the significant performance superiority of our proposal in achieving secure transmission as compared with the baselines.

\end{itemize}

The remainder of this article is organized as follows. In Section \ref{s1}, we introduce the related work. In Section \ref{s2}, we present the system model and problem formulation for the multi-ARIS assisted satellite system.  In Section \ref{s3}, we optimize the small-scale problems of transmission beamforming and passive beamforming. In Section \ref{s4}, we optimize the large-scale problems of ARIS association and deployment. Section \ref{s6} provides the simulation results to verify the effectiveness of the proposed algorithm design. Finally, Section \ref{s7} concludes this paper.

\textit{Notations}: Vectors and matrices are represented by bold letters. The transpose and conjugate transpose of a matrix $\boldsymbol{A}$ are represented by $\boldsymbol{A}^{\mathrm{T}}$ and $\boldsymbol{A}^{\dagger}$. For a square matrix $\boldsymbol{A}$, $\mathrm{Tr}(\boldsymbol{A})$ is the trace of $\boldsymbol{A}$. The expression $\mathrm{diag}(a)$ is the diagonal matrix with the elements of $a$, and also $\mathrm{diag}(\boldsymbol{A})$ returns a diagonal matrix with the elements of $\boldsymbol{A}$ on its main diagonal. For a complex scalar number $a$, $|a|$ is used for the absolute value of $a$ and $\Re(a)$ denotes the real part of a complex scalar $a$. For a matrix $\boldsymbol{A}$, $||\boldsymbol{A}||$ denotes the spectral norm of $\boldsymbol{A}$, $\mathrm{rank}(\boldsymbol{A})$ implies the rank of $\boldsymbol{A}$ and $\boldsymbol{A}\succeq0$ represents a positive semi-definite matrix $\boldsymbol{A}$. The expression $\mathcal{O}(\cdot)$ implies the big-O notation.

\section{Related Work}\label{s1}

Recently, physical layer security technology has garnered significant attention as a promising solution to enhance the security of satellite communications. Among various physical layer security enhancement strategies, relay cooperation technology has emerged as a promising approach to secure satellite communications. Existing research has suggested using relay nodes to ensure secure transmission in a hybrid satellite terrestrial relay network~\cite{a24} and mitigate eavesdropping by emitting artificial noise (AN) through a relay node~\cite{a25}. Meanwhile, some researchers utilize aerial relay nodes, which can be flexibly deployed to further enhance satellite communications security. In~\cite{a26}, the authors proposed the use of aerial relay nodes to introduce the air channel difference to enhance physical layer security. In~\cite{a27}, the authors employ a UAV as both relays and jammers, generating AN to disrupt eavesdroppers. Another promising strategy to enhance satellite communication security is using traditional terrestrial wireless communication techniques such as precoding and beamforming. In~\cite{a28}, the authors investigated the optimization of transmission beamforming to improve physical layer security performance in multi-beam satellite systems, thereby establishing a foundation for this emerging research area. In~\cite{a29}, the authors propose a deep reinforcement learning framework to jointly design UAV deployment and secure beamforming, thereby enabling secure communication for multiple users. Furthermore, in~\cite{R2}, the authors focused on leveraging the interference from the terrestrial base station to secure the satellite link, and investigated the secrecy-energy efficiency of hybrid beamforming schemes under the practical assumption of imperfect channel state information. However, the long propagation distance of satellite communications results in highly correlated channel conditions for the legitimate user and the nearby eavesdropper, making it difficult for conventional security techniques to achieve the desired effect. These limitations highlight the need for enhanced security approaches in satellite communications.

With RIS emerging as a promising solution for next-generation communication systems, several researchers have investigated the potential of RIS-assisted secure communications. These works focus on the joint design of active transmission and passive reflection beamforming, combined with advanced optimization techniques, to enhance the physical layer security of wireless communications. In~\cite{a32}, the authors use RIS to enhance physical layer security by jointly optimizing the transmission beamforming and passive beamforming of RIS. In~\cite{a33}, the authors propose closed-form solutions for optimizing the passive beamforming of RIS to maximize secrecy when eavesdropping channel information is unavailable. In~\cite{R3}, the authors employ a multi-functional RIS, integrating its capabilities with semantic communications to establish a robust anti-jamming system for complex aerial-ground networks. Meanwhile, to address the practical power supply issue for RIS, the authors in~\cite{R4} explored a self-sustaining, absorptive ground-based RIS that can concurrently harvest energy and enhance security, while offering flexible control over the amplitude and phase of reflected signals. Although these proposals provide inspiring and effective solutions to enhance the security of wireless systems, they are heavily challenged by the practical constraint that RISs usually need to be deployed at fixed locations. Due to the adaptable and on-demand deployment capabilities of ARIS, it presents significant potential to advance conventional security methods for better protection of information security~\cite{a36,a37}. In~\cite{a38}, the authors proposed a method to achieve anti-eavesdropping communication by jointly optimizing the transmit beamforming, phase shifts, and location of the ARIS. In~\cite{a39}, the authors introduced a robust anti-eavesdropping communication system using ARIS, optimized through deep reinforcement learning, to enhance system security performance. In~\cite{a40}, the authors addressed the issue of anti-jamming aerial-ground communication and investigated the design of passive beamforming and flying trajectories to mitigate the jamming attacks.
\begin{table}[!t]
    \centering % 表格居中
    \setlength{\tabcolsep}{4pt}
    \caption{Comparison of Our Work with Closely Related Studies}
    \label{tab:feature_comparison}
    \begin{tabular}{|l|c|c|c|c|c|c|}
        \hline
        \diagbox{Properties}{References} & This paper& [38]& [39]& [40]& [41]& [42]\\
        \hline
        Fixed RIS                     &            & \checkmark & \checkmark & \checkmark & \checkmark &            \\
        \hline
        Aerial RIS& \checkmark &            &            &            &            & \checkmark \\
        \hline
        Multibeam Satellite           &            \checkmark&            &            & \checkmark & \checkmark &            \\
        \hline
        Multi-RIS                     & \checkmark &            &            &            &            &            \\
        \hline
        Multi-Eavesdropper            & \checkmark &            &            &            &            &            \\
        \hline
        ARIS-Group Association        & \checkmark &            &            &            &            &            \\
        \hline
    \end{tabular}
\end{table}

Given the significant advantages of RIS in enhancing physical layer security in terrestrial networks, the potential of RIS applications has also been exploited in the physical layer security of satellite communications. In~\cite{a41}, the authors investigated the physical layer security of multi-beam satellite communications assisted by ground-based RIS. In~\cite{a42}, the author combined active RIS and passive RIS to assist satellite downlink communication systems for secure transmission. Compared to traditional passive RIS schemes, the system's confidentiality performance is significantly improved. In~\cite{a43}, the authors proposed jointly optimizing the RIS and artificial noise from the terrestrial network to effectively safeguard satellite downlink transmissions against potential eavesdropping. In~\cite{R5}, the authors proposed a hybrid secure transmission scheme for RIS-assisted multibeam satellite communications under imperfect Channel State Information. However, these related works mainly focused on traditional terrestrial RIS, which is limited by fixed position. In~\cite{R6}, the authors enhanced the physical layer security performance of a cognitive non-terrestrial network by deploying the RIS on a UAV and optimizing the UAV's trajectory and the RIS's reflection coefficients. From Table~\ref{tab:feature_comparison}, we observe that most of the existing works focus on using terrestrial RIS to assist satellite physical layer security. Therefore, in this work, we further investigate the problem of ARIS-assisted satellite communications, and we utilize multiple ARISs to achieve effective user coverage. This innovative setup paves the way for the dynamic application of ARIS in satellite communications and highlights the significant advantages of ARIS compared to traditional scenarios.

\section{System Model and Problem Formulation}\label{s2}
\begin{figure}[t!]
\centering
    \includegraphics[width=1\linewidth]{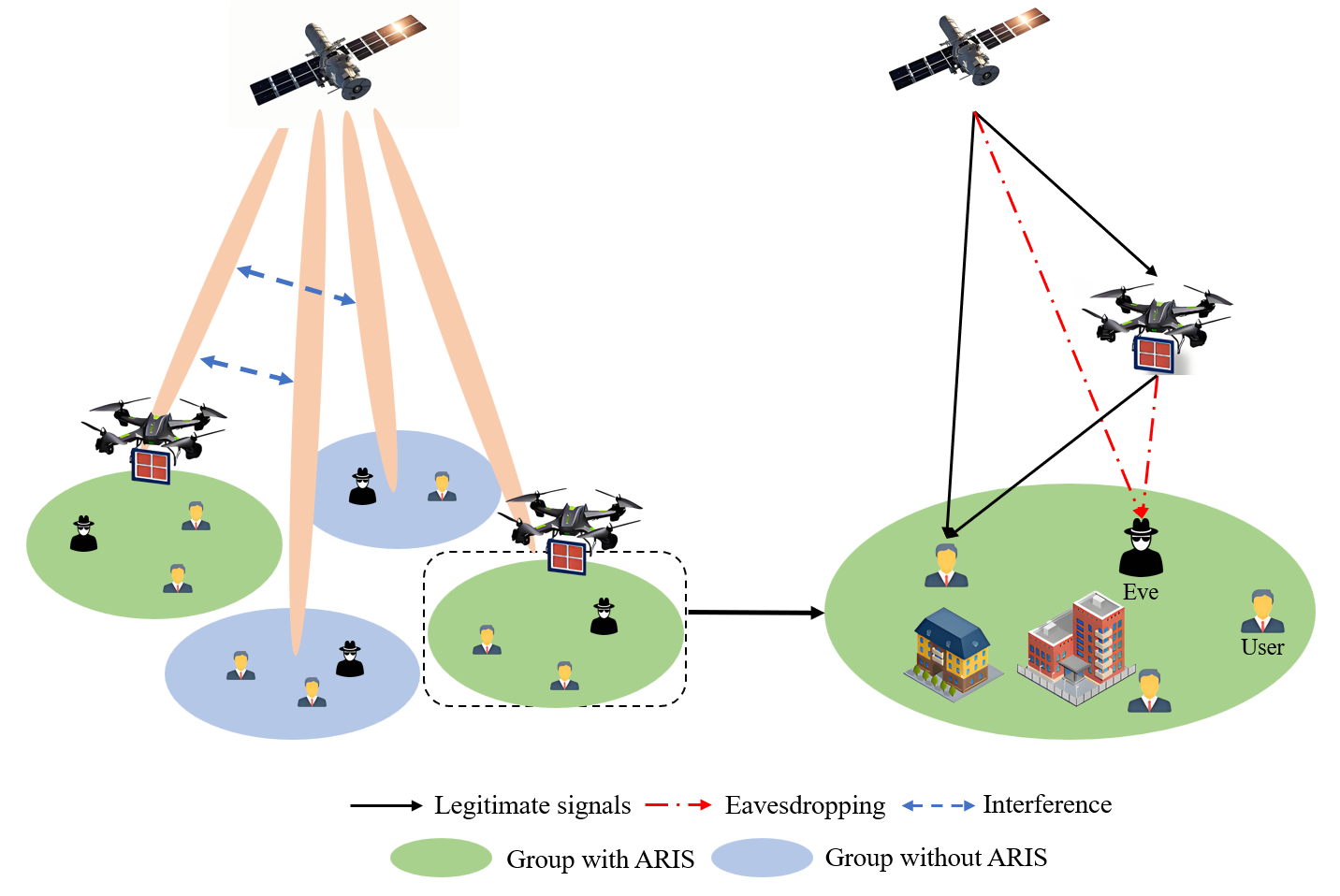}
    \caption{Multi-ARIS assisted satellite communication}
    \label{system}
\end{figure}
As shown in Fig. \ref{system}, we consider a multibeam satellite system consisting of $K$ groups, where the set of groups is denoted by $\mathcal{K}=\{1,\ldots,K\}$. Each beam is intended for a single group, where the users in the same group conduct multi-cast receptions. In the $k$-th group, the multi-cast transmissions have $M_k$ intended users along with $E_k$ unintended users, for which the transmissions need protection from the unintended ones. The sets of intended and unintended users are also termed as $\mathcal{M}_k=\{1,\ldots,M_k\}$ and $\mathcal{E}_k=\{1,\ldots,E_k\}$, where the collection of all ground users within the $k$-th group can be denoted by $\mathcal{I}_k=\mathcal{M}_k\cup\mathcal{E}_k$. Meanwhile, the multibeam satellite is equipped with $L$ antennas, and the ground user $i_k\in\mathcal{I}_k$ is equipped with one single antenna. The two-dimensional horizontal coordinates of the ground user $i_k\in\mathcal{I}_k$ is denoted by $\boldsymbol{\omega}_{i_k}=[x_{i_k},y_{i_k}]$, and it is located within an area defined by $\mathcal{X}\times\mathcal{Y}$ with $\mathcal{X}$ and $\mathcal{Y}$ denoting the range along x-axis and y-axis, respectively.

For the considered scenarios, the limited degrees of freedom in beamforming make it difficult to ensure the security in satellite communications. We propose to deploy $J$ aerial platforms, each equipped with an RIS\footnote{For the ARIS, we assume that the aerial platform is sufficiently stable that the determined location and phase shifts can be maintained while the platform is hovering.}. The set of AIRSs is denoted by $\mathcal{J}=\{1,\dots,J\}$. Given the constraints on cost, it is generally difficult to deploy an ARIS in each group. Therefore, we consider the optimal association strategy between the ARISs and the groups when the number of ARISs is fewer than that of groups. The association coefficients are defined as $\chi_{j,k}\in\{0, 1\}$, which show the association between the  $j$-th ARIS and the $k$-th groups. The altitude and horizontal coordinates of the $j$-th ARIS are denoted by $H_j$ and $\boldsymbol{q}_j=[x_j,y_j]$, respectively. Thus, the distance between the $j$-th ARIS and the ground user $i_k$ is $d_{j,i_k}=\sqrt{{\|\boldsymbol{q}_j-\boldsymbol{\omega}_{i_k}\|}^2+H^2}$. To enhance the reflection-based transmissions, we introduce the concept of subsurface that each ARIS comprises a number of adjacent elements inducing an identical phase shift to the incident signal~\cite{a11}. The ARIS consists of $N$ subsurfaces. For the $j$-th ARIS, denote the phase shifts of the subsurfaces as $\boldsymbol{\theta}_{j}=[\theta_{j,1},\theta_{j,2},\ldots,\theta_{j,N}]^{\mathrm{T}}$, then the reflection-coefficient matrix can be given as $\boldsymbol{{\Theta}}_{j}=\mathrm{diag}(\boldsymbol{\vartheta}_{j})$ with $\boldsymbol{\vartheta}_{j}=[e^{i\theta_{j,1}},e^{i\theta_{j,2}},\ldots,e^{i\theta_{j,N}}]^\mathrm{T}$. In addition, we assume the ARIS is cooperatively controlled via onboard processing with reliable backhaul links.

\subsection{Channel Model}
According to the signal propagation characteristics of the LEO satellite communications~\cite{a44,a45}, the channel from the satellite to the ground user $i_k$ is expressed as $\boldsymbol{h}_{i_k}=\bar{h}_{i_k}\cdot\boldsymbol{r}^{-\frac12}\odot\boldsymbol{b}^{\frac12}$, and $\boldsymbol{h}_{i_k}\in\mathbb{C}^{L\times1}$. The free space loss and phase of the channel are defined by  $\bar{h}_{i_k}=\frac\lambda{4\pi d_{s,i_k}}e^{-i\frac{2\pi}\lambda d_{s,i_k}}$, where ${\lambda}$ is the carrier wavelength, and $d_{s,i_k}$ is the distance between the satellite and the ground user. The rain attenuation $\boldsymbol{r} = \xi ^2 \boldsymbol{1} _{L\times 1}$ follows log-normal distribution with $\mathrm{ln}(\xi_\mathrm{dB})\sim$ $\mathcal{CN}(\mu,\sigma^{2})$, where $\mu$ and $\sigma$ are the mean value and variance which are related to the satellite communication frequency, polarization mode and the location of the user, respectively. Besides, the beam gain $\boldsymbol{b} = [ b_1,\ldots, b_L]^\mathrm{T}$ mainly depends on the satellite transmit antenna radiation pattern and the user location, where the gain is given by $b_l=G_{l,max}\left(\frac{J_1(u_l)}{2u_l}+36\frac{J_3(u_l)}{u_l^3}\right)^2$, with $u_l=2.07123\sin\varphi_{l}/\sin\left(\varphi_{\mathrm{3dB}}\right)_{l}$, where $G_{l,max}$ is the maximum beam gain for the $l$-th beam, $\varphi_\mathrm{3dB}$ is the angle corresponding to the 3 dB power loss from the beam center.  Additionally, $J_1$ and $J_3$ are the first-kind Bessel function with first order and third order, respectively.

The channel from the satellite to the $j$-th ARIS is denoted by $\boldsymbol{G}_j=[\boldsymbol{G}_{j,1},
\dots,\boldsymbol{G}_{j,N}]^{\mathrm{T}}$, and $\boldsymbol{G}_j\in\mathbb{C}^{N\times L}$. The channel gain between the satellite and the $n$-th subsurface of the $j$-th ARIS is given by $\boldsymbol{G}_{j,n}=\bar{G}_{j,n}\cdot\boldsymbol{r}^{-\frac12}\odot\boldsymbol{b}^{\frac12}$. Consider the ARIS forming a uniform linear array along x-axis, the free space loss and phase can be expressed as $\bar{G}_{j,n}=\frac\lambda{4\pi d_{s,j}}e^{-i\frac{2\pi}\lambda d_{s,j}}e^{-i\frac{2\pi d}\lambda(n-1)\phi_j}$, where $d_{s,j}$ is the distance between the satellite and the $j$-th ARIS, $d$ is the separation between two adjacent reflecting elements, $\phi_j$ denotes the cosine of angle of arrival. The rain attenuation is expressed by  $\boldsymbol{r} \in \mathbb{C} ^{1\times L}$, and the beam gain is denoted by $\boldsymbol{b} \in \mathbb{C} ^{1\times L }$.

The ARIS-to-ground transmission experiences both path loss and small-scale fading. The path loss from ARIS to the ground user is given by $C_L=L_0d_{j,i_k}^{-\beta}$, where $L_0$ denotes the path loss at the reference distance of 1 m, $d_{j,i_k}$ denotes the distance between the $j$-th ARIS and the ground user $i_k$ within the $k$-th group, $\beta$ is the path loss exponent for the wireless transmissions on the ground. As for small-scale fading, the Rician fading model is adopted. Therefore, the ARIS-to-ground channel is represented as $\boldsymbol{g}_{j,i_k}=\sqrt{C_L}(\sqrt{\frac\rho{\rho+1}}\boldsymbol{g}_{j,i_k}^\mathrm{LoS}+\sqrt{\frac1{\rho+1}}\boldsymbol{g}_{j,i_k}^\mathrm{NLoS})$,
where $\rho$ is the Rician factor, $\boldsymbol{g}_{j,i_k}^\mathrm{LoS}=e^{-i\frac{2\pi}{\lambda}d_{j,i_k}}[e^{-i\frac{2\pi d}{\lambda}(n-1)\phi_{j,i_k}}]_{n\in\mathcal{N}}$ denotes the deterministic LoS component which is related to the angle of departure $\phi_{j,i_k}=\frac{x_j-x_{i_k}}{d_{j,i_k}}$ and $\boldsymbol{g}_{j,i_k}^\mathrm{NLoS}$ denotes the non-LoS component modeled as Rayleigh fading with $\boldsymbol{g}_{j,i_k}^\mathrm{NLoS}\sim\mathcal{CN}(0,1)$. In addition, while we assume perfect CSI in this paper, the model can be extended to scenarios with imperfect CSI by leveraging the S-procedure and generalized sign-definiteness conditions.

\subsection{Signal Model}
For the considered system model, the received direct and reflected signals at the ground user $i_k$ are given as
\begin{equation}
\begin{aligned}\label{eq1}
y_{i_k}=\left(\boldsymbol{h}^{\dagger}_{i_k}+\sum\limits_{j=1}^J\chi_{j,k}\boldsymbol{g}^{\dagger}_{j,i_k}\boldsymbol{\Theta}_{j} \boldsymbol{G}_{j}\right)\sum_{l=1}^K\boldsymbol{w}_{l}s_{l}+n_{i_k},\\
\hfill  \forall i_k\in\mathcal{I}_k, \forall k\in\mathcal{K},
\end{aligned}    
\end{equation}
where $\boldsymbol{w}_{l}\in\mathbb{C}^{L\times 1}$ denotes the beamforming vector for the $l$-th group, along with the transmitted symbol $s_{l}\sim\mathcal{CN}(0,1)$, and $n_{i_k}$ is the white Gaussian noise at the ground user $i_k\in\mathcal{I}_k$. Additionally, the satellite beamforming is constrained by the maximum satellite power $P_T$, which is expressed by $\sum\limits_{k=1}^K\|\boldsymbol{w}_k\|^2\leq P_T$. Then, we obtain the signal-to-interference-plus-noise ratios (SINRs) of ground user $i_k$ as
\begin{equation}\label{eq4}
\begin{aligned}
\gamma_{i_k}=\frac{\left|\left(\boldsymbol{h}^{\dagger}_{i_k}+\sum\limits_{j=1}^J\chi_{j,k}\boldsymbol{g}^{\dagger}_{j,i_k}\boldsymbol{\Theta}_{j} \boldsymbol{G}_{j}\right){\boldsymbol{w}}_{k}\right|^{2}}{\sum\limits_{l=1,l\neq k}^K\left|\left(\boldsymbol{h}^{\dagger}_{i_k}+\sum\limits_{j=1}^J\chi_{j,k}\boldsymbol{g}^{\dagger}_{j,i_k}\boldsymbol{\Theta}_{j} \boldsymbol{G}_{j}\right){\boldsymbol{w}}_{l}\right|^{2}+\sigma_{i_k}^{2}},
\\ \hfill  \forall i_k\in\mathcal{I}_k, \forall k\in\mathcal{K}.
\end{aligned}
\end{equation}
The $\sigma_{i_k}^2=k_B B T_{i_k}$ is the noise power, where $k_B=1.380649\times10^{-23}J/K$ is the Boltzmann constant, $B$ is the operating bandwidth, and $T_{i_k}$ is the noise temperature. Then, the achievable rates of all ground users, including the rates of legitimate users $R_{m_k}$ and the rates of eavesdroppers $R_{e_k}$, can be written as
\begin{equation}\label{eq29}
    R_{i_k}=\log(1+\gamma_{i_k}),\quad\forall i_k\in\mathcal{I}_k,\forall k\in\mathcal{K}.
\end{equation}
\subsection{Proposed Solution}
To ensure the security of satellite communication, we employ multiple ARISs to enhance legitimate transmission while mitigating the information leakage to unintended parties. Therefore, the problem is formulated to enhance the sum rate of all groups while limiting the eavesdroppers' rate by jointly optimizing the transmission beamforming, the passive beamforming of ARISs, the ARIS association, and the deployment of the ARISs. The optimization of multigroup multicast sum rate maximization seeks to enhance the minimum rate within each group while maximizing the aggregate rates across all groups. Thus, the problem is formulated as 
\begin{subequations}\label{eq5}
\begin{align}
\max_{\boldsymbol{w}_{k},\boldsymbol{\theta}_j,\boldsymbol{q}_j,
\chi_{j,k}}\quad&\sum_{k=1}^{K} \min_{m_k\in\mathcal{M}_k}R_{m_k}\label{eq6}\\
\mathrm{s.t.}\quad&R_{e_k}\leq\Upsilon_k,\quad\forall e_k\in \mathcal{E}_k, \forall k\in\mathcal{K},\label{con1}\\
&\sum_{k=1}^K\|\boldsymbol{w}_k\|^2\leq P_T,\label{con2}\\
&\theta_{j,n}\in[0,2\pi),\quad\forall j\in\mathcal{J}, \forall n\in\mathcal{N},\label{con3}\\
&\boldsymbol{q}_j\in\mathcal{X}\times\mathcal{Y},\quad\forall j\in\mathcal{J},\label{all1}\\
&\chi_{j,k}\in\{0,1\},\quad\forall j\in\mathcal{J}, \forall k\in\mathcal{K},\label{all2}\\
&\sum\limits_{k=1}^K\chi_{j,k}=1,\quad\forall j\in\mathcal{J},\label{all3}\\
&\sum\limits_{j=1}^J\chi_{j,k}\leq1,\quad\forall k\in\mathcal{K},\label{all4}
\end{align}
\end{subequations}
where $\Upsilon_k$ in (\ref{con1}) is the introduced threshold of wiretap rate for eavesdroppers $E_k$, and (\ref{con3}) is the unit-modulus constraint on ARISs' phase shifters. As shown in (\ref{all3}) and (\ref{all4}), it is assumed that each group is associated with only one ARIS. Notably, our model can be conveniently extended to address energy or latency issues by reformulating the rate constraint into a data-volume constraint that reflects practical application requirements. In addition, inter-group fairness can be conveniently achieved by extending the objective function with Jain’s Fairness Index.

The formulated problem is rather complicated with threefold difficulties. First, the objective of the problem is non-convex due to the minimization operation. Second, we can readily see that the objective is a complicated function with respect to the optimization variables including the transmission beamforming, passive beamforming of ARISs, ARIS association, and ARIS deployment. Third, the $\chi_{j,k}$ is a binary variable and other variables are continuous, the problem (\ref{eq5}) is a mixed discrete-continuous problem.

In order to tackle the non-continuity reduced from the minimization operation in the objective, we propose to introduce a variable $\Omega_k$. Then, the problem is reformulated as
\begin{subequations}\label{con0}
\begin{align}
\max_{\boldsymbol{w}_{k},\boldsymbol{\theta}_j,\boldsymbol{q}_j,
\chi_{j,k},\Omega_k}\quad&\sum_{k=1}^{K} \Omega_k \\
\mathrm{s.t.}\quad&R_{m_k}\geq \Omega_k,\quad\forall m_k\in \mathcal{M}_k,\forall k\in\mathcal{K},\label{con4}\\
&(\ref{con1}), (\ref{con2}), (\ref{con3}), (\ref{all1}),(\ref{all2}), (\ref{all3}), (\ref{all4}),\nonumber
\end{align}
\end{subequations}
where the introduction of the variable $\Omega_k$ results in the new constraint (\ref{con4}). 

For the problem in (\ref{con0}), we can see that it is still rather complicated and non-convex. This complexity can primarily be attributed to constraints (\ref{con1}) and (\ref{con4}) are non-convex with respect to $\{\boldsymbol{w}_{k}\}_{k=1}^K$ and $\{\boldsymbol{\theta}_j\}_{j=1}^J$, and  (\ref{con3}) is nonlinear equality constraint. To solve the problem effectively, we propose a BCD framework to solve the problem sub-optimally. Due to the non-convex objective and constraints of the original problem (\ref{eq5}) and the tight coupling of all optimization variables, a direct joint optimization attempt would result in slow convergence and a high risk of the algorithm getting trapped in a suboptimal solution. Therefore, we partition the problem into large-scale and small-scale subproblems based on the different physical scales at which the variables affect system performance.

\section{Transmission and Reflection Optimization}\label{s3}
In this section, we will optimize the small-scale subproblems of transmission beamforming and passive beamforming. We reformulate each problem as a SDP problem and solve each by relaxing the rank-one constraint.
\subsection{Transmission Beamforming}
With the fixed passive beamforming, ARIS association, and  ARIS deployment, the 
 subproblem of transmission beamforming can be expressed as
\begin{align}\label{p1}
\max_{\boldsymbol{w}_{k},\Omega_k} \quad&\sum_{k=1}^{K} \Omega_k \\
\mathrm{s.t.}\quad&(\ref{con1}), (\ref{con2}),(\ref{con4})\nonumber.
\end{align}
Due to the objective function is now a linear function, the non-convexity lies in the constraints (\ref{con1}) and (\ref{con4}). To simplify the notations, we define
\begin{equation}
\begin{aligned}
\boldsymbol{\Xi}_{i_k}=&\Bigg(\boldsymbol{h}^{\dagger}_{i_k}+\sum\limits_{j=1}^J\chi_{j,k}\boldsymbol{g}^{\dagger}_{j,i_k}\boldsymbol{\Theta}_{j} \boldsymbol{G}_{j}\Bigg)\Bigg(\boldsymbol{h}_{i_k}^{\dagger}\\
&+\sum\limits_{j=1}^J\chi_{j,k}\boldsymbol{g}^{\dagger}_{j,i_k}\boldsymbol{\Theta}_{j} \boldsymbol{G}_{j}\Bigg)^{\dagger},\quad\forall i_k\in\mathcal{I}_k, \forall k\in\mathcal{K}.
\end{aligned}    
\end{equation}
Then, by defining $\boldsymbol{W}_k=\boldsymbol{w}_k{\boldsymbol{w}_k}^{\dagger}$, and the equivalent power constraint $\sum\limits_{k=1}^K\mathrm{Tr}(\boldsymbol{W}_k)\leq P_T$, the rates of legitimate users $R_{m_k}$ and the rates of eavesdroppers $R_{e_k}$ can be reformulated as 
\begin{equation}\label{eq8}
\begin{split}
R_{i_k}=&\log\left(\frac{\sum\limits_{l=1}^{K}\mathrm{Tr}(\boldsymbol{\Xi}_{i_k}\boldsymbol{W}_{l})+\sigma_{i_k}^{2}}{\sum\limits_{l=1,l\neq k}^K\mathrm{Tr}(\boldsymbol{\Xi}_{i_k}\boldsymbol{W}_l)+\sigma_{i_k}^{2}}\right)\\
=&\log\left(\sum_{l=1}^{K} \mathrm{Tr} (\boldsymbol{\Xi}_{i_k}\boldsymbol{W}_{l})+\sigma_{i_k}^{2}\right)\\
&-\log\left(\sum_{l=1,l\neq k}^K \mathrm{Tr} (\boldsymbol{\Xi}_{i_k}\boldsymbol{W}_{l})+\sigma_{i_k}^{2}\right),\\
&\forall i_k\in\mathcal{I}_k, \forall k\in\mathcal{K}.
\end{split}
\end{equation}

For the non-convex constraints (\ref{con1}) and (\ref{con4}), based on [45, Lemma 1], we can replace this term with the following equalities by introducing auxiliary variables $\xi_{m_k}$ and $\xi_{e_k}$,
\begin{equation}\label{eq10}
    \begin{aligned}&\log(\xi_{m_k})-\xi_{m_k}\left(\sum_{l=1,l\neq k}^{K} \mathrm{Tr} (\boldsymbol{\Xi}_{m_k}\boldsymbol{W}_{l})+\sigma_{m_k}^{2}\right)+1\\
&+\log\left(\sum_{l=1}^{K} \mathrm{Tr} (\boldsymbol{\Xi}_{m_k}\boldsymbol{W}_{l})+\sigma_{m_k}^{2}\right)>\Omega_{k},\\
&\forall m_k\in\mathcal{I}_k, \forall k\in\mathcal{K},
    \end{aligned}
\end{equation}
and
\begin{equation}\label{eq11}
 \begin{aligned}
&\log(\xi_{e_k})+\xi_{e_k}\left(\sum_{l=1}^{K} \mathrm{Tr} (\boldsymbol{\Xi}_{e_k}\boldsymbol{W}_{l})+\sigma_{e_k}^{2}\right)-1\\
&-\log\left(\sum_{l=1,l\neq k}^{K} \mathrm{Tr} (\boldsymbol{\Xi}_{e_k}\boldsymbol{W}_{l})+\sigma_{e_k}^{2}\right)\leq\Upsilon_k,\\
&\forall e_k\in\mathcal{E}_k, \forall k\in\mathcal{K}.
 \end{aligned}
\end{equation}
Then, the transmission beamforming optimization problem becomes
\begin{subequations}\label{eq13}
\begin{align} 
\max_{\boldsymbol{W}_{k},\xi_{m_k},\xi_{e_k},\Omega_k}\quad&\sum_{k=1}^{K}\Omega_k \\
\mathrm{s.t.}
\quad&\xi_{m_k}>0,\quad\forall m_k\in \mathcal{M}_k, \forall k\in\mathcal{K},\\ &\xi_{e_k}>0,\quad\forall e_k\in \mathcal{E}_k, \forall k\in\mathcal{K},\\
&\boldsymbol{W}_k\succeq0, \mathrm{rank}\left(\boldsymbol{W}_k\right)=1,\quad\forall k\in\mathcal{K},\label{n3}\\
&\sum_{k=1}^{K}\mathrm{Tr}(\boldsymbol{W}_{k})\leq P_{T},\\
&(\ref{eq10}),(\ref{eq11})\nonumber,
\end{align}
\end{subequations}
where the use of optimization variables $\{\boldsymbol{W}_k\}_{k=1}^K$ rather than $\{\boldsymbol{w}_k\}_{k=1}^K$ leads to the rank-1 constraint given in equation (\ref{n3}). For this problem, we can similarly optimize the auxiliaries and transmission beamforming alternatively, where the optimized $\xi_{m_k}$ and $\xi_{e_k}$ are obtained as
\begin{equation}
    \xi_{m_k}=\frac{1}{\sum\limits_{l=1,l\neq k}^K\mathrm{Tr}(\boldsymbol{\Xi}_{m_k}\boldsymbol{W}_{l})+\sigma_{m_k}^{2}},\quad\forall m_k\in \mathcal{M}_k, \forall k\in\mathcal{K},
\end{equation}
and
\begin{equation}
    \xi_{e_k}=\frac{1}{\sum\limits_{l=1}^{K}\mathrm{Tr} (\boldsymbol{\Xi}_{e_k}\boldsymbol{W}_{l})+\sigma_{e_k}^{2}},\quad\forall e_k\in \mathcal{E}_k, \forall k\in\mathcal{K},
\end{equation}
respectively. Meanwhile, for optimization variable $\{\boldsymbol{W}_{k}\}_{k=1}^K$, by applying the semidefinite relaxation and temporarily ignoring the rank-1 constraint, it becomes convex semidefinite programming and can be conveniently solved by standard toolbox such as CVX. Finally, the solution to the transmission beamforming optimization problem is obtained by iteratively optimizing $\{\xi_{m_k}\}_{k=1}^K$, $\{\xi_{e_k}\}_{k=1}^K$ and $\{\boldsymbol{W}_{k}\}_{k=1}^K$ in (\ref{eq13}). Finally, the transmission beamforming vector $\boldsymbol{w}_{k}$ can be obtained by eigenvalue decomposition of $\{\boldsymbol{W}_{k}\}_{k=1}^K$, which is further processed with Gaussian Randomization if $\{\boldsymbol{W}_{k}\}_{k=1}^K$ fails the rank-1 constraint.

\subsection{Passive Beamforming of ARISs}
With the fixed transmission beamforming, ARIS association and ARIS deployment, the subproblem of passive beamforming can be expressed as
\begin{align}\label{p2}
\max_{\boldsymbol{\theta}_{j},\Omega_k} \quad&\sum_{k=1}^{K} \Omega_k \\
\mathrm{s.t.}\quad&(\ref{con1}), (\ref{con3}), (\ref{con4})\nonumber.
\end{align}
Furthermore, in the considered system, there is a significant distance between different groups. Therefore, for a given group, we only consider the reflection links from the ARIS associated with this group, while the reflection links from ARISs in other groups are excluded. Meanwhile, the impact of ARISs on unassociated groups will also be ignored.

For notation simplicity, we define
\begin{equation}
 \begin{aligned}\label{eq2}
 &\boldsymbol{\boldsymbol{v}}_{j}=[1,\boldsymbol{\vartheta}_{j}],\quad\forall j\in\mathcal{J},\\ &\boldsymbol{H}_{i_k}=[\boldsymbol{h}_{i_k}^{\dagger};\sum\limits_{j=1}^J\chi_{j,k}\mathrm{diag}(\boldsymbol{g}_{j,i_k}^{\dagger})\boldsymbol{G}_{j}],\quad\forall i_k\in\mathcal{I}_k, \forall k\in\mathcal{K}.
\end{aligned}   
\end{equation}
Then, the SINRs of ground user $i_k$ can be rewritten in a more condensed form as
\begin{equation}\label{eq3}
\gamma_{i_k}=\frac{|\boldsymbol{\boldsymbol{v}}_{j}\boldsymbol{H}_{i_k}\boldsymbol{w}_{k}|^{2}}{\sum\limits_{l=1,l\neq k}^K|\boldsymbol{\boldsymbol{v}}_{j}\boldsymbol{H}_{i_k}\boldsymbol{w}_{l}|^{2}+\sigma_{i_k}^{2}},\quad\forall i_k\in\mathcal{I}_k,\forall k\in\mathcal{K}.
\end{equation}
To facilitate the discussions, we define
\begin{equation}
 \begin{aligned}
    &\boldsymbol{\Lambda}_{i_k,l}= \boldsymbol{H}_{i_k}\boldsymbol{w}_l\boldsymbol{w}_l^{\dagger}\boldsymbol{H}_{i_k}^{\dagger},\quad\forall i_k\in\mathcal{I}_k, \forall k\in\mathcal{K}.
\end{aligned}   
\end{equation}
Then, by defining $\boldsymbol{V}_{j}=\boldsymbol{v}_{j}^{\dagger}\boldsymbol{v}_{j}$, the rates of legitimate users $R_{m_k}$ and the rates of eavesdroppers $R_{e_k}$ can be reformulated as
\begin{equation}\small
 \begin{aligned}
R_{i_k}=&\log\left(\frac{\sum\limits_{l=1}^{K} \mathrm{Tr}(\boldsymbol{\Lambda}_{i_k,l}\boldsymbol{V}_{j})+\sigma_{i_k}^{2}}{\sum\limits_{l=1,l\neq k}^{K} \mathrm{Tr}(\boldsymbol{\Lambda}_{i_k,l}\boldsymbol{V}_{j})+\sigma_{i_k}^{2}}\right)\\
=&\log\left(\sum_{l=1}^{K} \mathrm{Tr}(\boldsymbol{\Lambda}_{i_k,l}\boldsymbol{V}_{j})+\sigma_{i_k}^{2}\right)\\
&-\log\left(\sum_{l=1,l\neq k}^{K} \mathrm{Tr}(\boldsymbol{\Lambda}_{i_k,l}\boldsymbol{V}_{j})+\sigma_{i_k}^{2}\right),\\&\quad\forall i_k\in\mathcal{I}_k,  \forall k\in\mathcal{K}.
\end{aligned}   
\end{equation}

Assuming that the reflection coefficient matrices of the ARISs in the other groups are fixed, we focus on the $j$-th ARIS which is associated with the $k$-th group. Therefore, the problem in (\ref{p2}) can be further decomposed into the problem of the $j$-th ARIS, which can be reformulated as
\begin{subequations}
\begin{align}
\max_{\boldsymbol{\theta}_j,\Omega_k} \quad&\Omega_k \\
\mathrm{s.t.}\quad&R_{m_k}\geq \Omega_k,\quad\forall m_k\in \mathcal{M}_k,\label{j1}\\
&R_{e_k}\leq\Upsilon_k,\quad\forall e_k\in \mathcal{E}_k, \label{j2}\\
&\theta_{j,n}\in[0,2\pi),\quad\forall n\in\mathcal{N}.
\end{align}
\end{subequations}
For the passive beamforming optimization problem, the constraints (\ref{j1}) and (\ref{j2}) are non-convex with respect to passive beamforming strategy. By employing the same methodology as presented in equations (\ref{eq10}) and (\ref{eq11}), the introduction of auxiliary variables $\mu_{m_k}$ and $\mu_{e_k}$ results in the following equations
\begin{equation}\label{eq14}
    \begin{aligned}
  &\log(\mu_{m_k})-\mu_{m_k}\left(\sum_{l=1,l\neq k}^{K} \mathrm{Tr} (\boldsymbol{\Lambda}_{m_k,l}\boldsymbol{V}_{j})+\sigma_{m_k}^{2}\right)+1\\
&+\log\left(\sum_{l=1}^{K} \mathrm{Tr} (\boldsymbol{\Lambda}_{m_k,l}\boldsymbol{V}_{j})+\sigma_{m_k}^{2}\right)>\Omega_k,\quad\forall m_k\in\mathcal{M}_k,\\
    \end{aligned}
\end{equation}
and
\begin{equation}\label{eq15}
\begin{aligned}
&\log(\mu_{e_k})+\mu_{e_k}\left(\sum_{l=1}^{K} \mathrm{Tr} (\boldsymbol{\Lambda}_{e_k,l}\boldsymbol{V}_{j})+\sigma_{e_k}^{2}\right)-1\\
&-\log\left(\sum_{l=1,l\neq k}^{K} \mathrm{Tr} (\boldsymbol{\Lambda}_{e_k,l}\boldsymbol{V}_{j})+\sigma_{e_k}^{2}\right)\leq\Upsilon_k,\quad\forall e_k\in\mathcal{E}_k.
\end{aligned}
\end{equation}

With the results in (\ref{eq14}) and (\ref{eq15}), the transmission beamforming optimization problem becomes
\begin{subequations}\label{eq16}
\begin{align}
\max_{\boldsymbol{V}_{j},\mu_{m_k},\mu_{e_k},\Omega_k}\quad& \Omega_k \\
\mathrm{s.t.}
\quad&\mu_{m_k}>0,\quad\forall m_k\in \mathcal{M}_k,\\
&\mu_{e_k}>0,\quad\forall e_k\in \mathcal{E}_k,\\
&\boldsymbol{V}_j\succeq0,  \mathrm{rank}(\boldsymbol{V}_{j})=1,\label{n1}\\
&(\boldsymbol{V}_{j})_{n,n} = 1,\quad\forall n\in \mathcal{N},\label{n2}\\
&(\ref{eq14}), (\ref{eq15}),\nonumber
\end{align}
\end{subequations}
where the employ of the optimization variables $\boldsymbol{V}_j$ rather than $\boldsymbol{v}_j$ leads to the rank-1 constraint given in equation (\ref{n1}). Also, The original constraint in (\ref{con3}) is reinterpreted as (\ref{n2}). For the problem in (\ref{eq16}), we can similarly optimize the auxiliaries and phase shift alternatively, where the optimized $\mu_{m_k}$ and $\mu_{e_k}$ are obtained as
\begin{equation}
    \mu_{m_k}=\frac{1}{\sum\limits_{l=1,l\neq k}^K\mathrm{Tr}(\boldsymbol{\Lambda}_{m_k,l}\boldsymbol{V}_{j})+\sigma_{m_k}^{2}},\quad\forall m_k\in \mathcal{M}_k,
\end{equation}
and
\begin{equation}
    \mu_{e_k}=\frac{1}{\sum\limits_{l=1}^{K}\mathrm{Tr} (\boldsymbol{\Lambda}_{e_k,l}\boldsymbol{V}_{j})+\sigma_{e_k}^{2}},\quad\forall e_k\in\mathcal{E}_k.
\end{equation}

For the phase shift, we can resort to semidefinite relaxation and ignore the rank-1 constraint and solve the problem with toolboxes like CVX. Then, the Gaussian randomization can be applied if the rank-1 constraint is not satisfied at the obtained optimum. Moreover, as the first element of  $\boldsymbol{\boldsymbol{v}}_{j}$ is always one, the phase shift vector is finally arrived according to $\vartheta_{j,n}=\exp(i\angle(v_{j,n+1}/v_{j,1})),\forall n\in\mathcal{N}$. Notably, our model can be conveniently extended to support discrete phase shift design by restricting each phase shift $\vartheta_{j,n}$ to a discrete set, and subsequently quantizing the continuous solution to the nearest feasible point in the discrete set.

\section{Association and Deployment Optimization}\label{s4}
In this subsection, we optimize the large-scale subproblems of ARIS association and deployment. For the ARIS association, we transform the mixed discrete-continuous problem by relaxing the binary variables and optimize it using SCA method. For the ARIS deployment, we also solve it using the same method.
\subsection{ARIS Association}\label{ap1}
In this subsection, we present the ARIS association scheme, which aims to design a stable association method that maximizes the sum rate of the considered system. With the fixed transmission beamforming, passive beamforming and ARIS deployment, the subproblem of ARIS association can be expressed as
\begin{align}\label{p3}
    \max_{\chi_{j,k},\Omega_k}\quad&\sum\limits_{k=1}^K\Omega_k\\
\mathrm{s.t.}\quad &(\ref{con1}), (\ref{all2}), (\ref{all3}), (\ref{all4}),(\ref{con4})\nonumber.
\end{align}
This subproblem is challenging to solve directly due to the complexity of the constraints (\ref{con1}) and (\ref{con4}) with respect to the variable $\chi_{j,k}$. By introducing variables $\boldsymbol{\chi}_k=[1,\chi_{1,k},\dots,\chi_{J,k}]$, $\boldsymbol{\Gamma}_{i_k}=[h_{i_k}^{\dagger};\boldsymbol{g}_{1,i_k}^{\dagger}\boldsymbol{\Theta}_1\boldsymbol{G}_1;\dots;\boldsymbol{g}_{J,i_k}^{\dagger}\boldsymbol{\Theta}_J\boldsymbol{G}_J]$ and $\boldsymbol{C}_{i_k,l}=\boldsymbol{\Gamma}_{i_k}\boldsymbol{w}_l\boldsymbol{w}_l^{\dagger}\boldsymbol{\Gamma_{i_k}}^{\dagger}$. The rate of ground user $i_k$ within the $k$-th group can be rewritten as
\begin{align}\label{u1}
    R_{i_k} =&\log\left(\sum\limits_{l=1}^K\boldsymbol{\chi}_k\boldsymbol{C}_{i_k,l}\boldsymbol{\chi}_k^{\mathrm{T}}+\sigma_{i_k}^2\right)\nonumber\\
&-\log\left(\sum\limits_{l=1,l\neq k}^K\boldsymbol{\chi}_k\boldsymbol{C}_{i_k,l}\boldsymbol{\chi}_k^{\mathrm{T}}+\sigma_{i_k}^2\right),\\
&\forall i_k\in\mathcal{I}_k, \forall k\in\mathcal{K}.\nonumber
\end{align}

Since this optimization problem incorporates binary variables, we first reformulate the binary constraint (\ref{all2}) as an equivalent equality constraint
\begin{align}\label{all9}
    \chi_{j,k}-(\chi_{j,k})^2=0,\quad\forall j\in\mathcal{J}, \forall k\in\mathcal{K}.
\end{align}
However, the equality constraint in equation (\ref{all9}) is only satisfied for binary variables. Then, we apply the penalty-based optimization algorithm to solve this problem. By relaxing the binary variables to continuous variables and incorporating the equality constraint (\ref{all9}) as a penalty term in the objective, the problem in (\ref{p3}) can be written as
\begin{subequations}\label{u2}
\begin{align}
\max_{\chi_{j,k},\Omega_k,\Delta_1}\quad&\sum\limits_{k=1}^K\left(\Omega_k-\tau\sum\limits_{j=1}^J(\chi_{j,k}-(\chi_{j,k})^2)\right)\label{all10}\\
\mathrm{s.t.}\quad&0\leq\chi_{j,k}\leq1,\quad\forall j\in\mathcal{J}, \forall k\in\mathcal{K},\label{all12}\\
&(\ref{con1}), (\ref{all3}), (\ref{all4}),(\ref{con4})\nonumber,
\end{align}
\end{subequations}
where $\tau>0$ is a penalty factor which penalizes the objective function.

The problem in (\ref{u2}) exhibits inherent non-convexity due to structural complexities in both its objective function and constraints. The objective function incorporates concave penalty terms. Meanwhile, the non-convex constraints primarily from the composition of concave logarithmic functions with quadratic terms and the concave-difference structure formed by subtracting two logarithmic functions. To solve the non-convexity of the constraints, we introduce a slack variable set $\Delta_1=\{\bar{\psi}_{i_k},\hat{\psi}_{i_k}\}$. Then, the constraints (\ref{con1}) and (\ref{con4})  can be reformulated as
\begin{subequations}\label{all5}
\begin{align}
&\bar{\psi}_{m_k}\leq\sum\limits_{l=1}^K\boldsymbol{\chi}_k\boldsymbol{C}_{m_k,l}\boldsymbol{\chi}_k^{\mathrm{T}}+\sigma_{m_k}^2,\label{new1}\\
&\hat{\psi}_{m_k}\geq\sum\limits_{l=1,l\neq k}^K\boldsymbol{\chi}_k\boldsymbol{C}_{m_k,l}\boldsymbol{\chi}_k^{\mathrm{T}}+\sigma_{m_k}^2,\label{u3}\\
&\log(\bar{\psi}_{m_k})-\log(\hat{\psi}_{m_k})\geq \Omega_k,\quad\forall m_k\in\mathcal{M}_k, \forall k\in\mathcal{K},\label{all7}
\end{align}
\end{subequations}
and
\begin{subequations}\label{all6}
\begin{align}
&\bar{\psi}_{e_k}\geq\sum\limits_{l=1}^K\boldsymbol{\chi}_k\boldsymbol{C}_{e_k,l}\boldsymbol{\chi}_k^{\mathrm{T}}+\sigma_{e_k}^2,\label{u4}\\
&\hat{\psi}_{e_k}\leq\sum\limits_{l=1,l\neq k}^K\boldsymbol{\chi}_k\boldsymbol{C}_{e_k,l}\boldsymbol{\chi}_k^{\mathrm{T}}+\sigma_{e_k}^2,\label{new2}\\
&\log(\bar{\psi}_{e_k})-\log(\hat{\psi}_{e_k})\leq \Upsilon_k,\quad\forall e_k\in\mathcal{E}_k, \forall k\in\mathcal{K}.\label{all8}
\end{align}
\end{subequations}
However, it is obvious that the constraints (\ref{new1}), (\ref{all7}),  (\ref{new2}) and (\ref{all8}) are non-convex. To address this issue, the SCA method is adopted to handle this problem. By replacing these non-convex terms with their FTSs, the constraints are rewritten as  
\begin{subequations}\label{u5}
\begin{align}
&\bar{\psi}_{m_k}\leq\sum\limits_{l=1}^K(\boldsymbol{\chi}_k\boldsymbol{C}_{m_k,l}\boldsymbol{\chi}_k^{\mathrm{T}})^{\mathrm{lb}}+\sigma_{m_k}^2,\\
&\log(\bar{\psi}_{m_k})-(\log(\hat{\psi}_{m_k}))^{\mathrm{ub}}\geq \Omega_k,\\
&\hat{\psi}_{e_k}\leq\sum\limits_{l=1,l\neq k}^K(\boldsymbol{\chi}_k\boldsymbol{C}_{e_k,l}\boldsymbol{\chi}_k^{\mathrm{T}})^{\mathrm{lb}}+\sigma_{e_k}^2,\\
&(\log(\bar{\psi}_{e_k}))^{\mathrm{ub}}-\log(\hat{\psi}_{e_k})\leq \Upsilon_k,
    \end{align}
\end{subequations}
respectively. Meanwhile, we can also apply SCA to handle the objective function. The FTSs of these terms are given in the Appendix. The feasible transformation sets (FTSs) of these terms are given in Appendix. Ultimately, the original problem (\ref{p3}) is transformed into the following convex form:
\begin{subequations}\label{all18}
\begin{align}
\max_{\chi_{j,k},\Omega_k,\Delta_1}&\quad\sum\limits_{k=1}^K\left(\Omega_k-\tau\sum\limits_{j=1}^J(\chi_{j,k}-(\chi_{j,k})^2)^{\mathrm{ub}}\right)\\
\mathrm{s.t.}&\quad (\ref{all3}), (\ref{all4}), (\ref{all12}), (\ref{u3}), (\ref{u4}), (\ref{u5}),\nonumber
\end{align}
\end{subequations}
where the constraints in (\ref{u3}), (\ref{u4}), and (\ref{u5}) are the convexified counterparts of (\ref{con1}) and (\ref{con4}). The problem in (\ref{all18}), as an approximated version of (\ref{p3}), incorporates a convex objective function and all convex constraints. Then, the original deployment optimization can be tackled by by using the SCA method.

\subsection{ARIS Deployment}\label{ap2}
With fixed transmission beamforming, passive beamforming and ARIS association, the subproblem of ARIS deployment can be expressed as
\begin{align}\label{p4}
\max_{\boldsymbol{q}_j,\Omega_k}\quad&\sum\limits_{k=1}^K\Omega_k\\
\mathrm{s.t.}\quad&(\ref{con1}),(\ref{all1}),(\ref{con4})\nonumber.
\end{align}
Due to the large distance between the satellite and the ARIS, the distance between the ARIS and the ground user becomes the most important factor affecting the quality of the reflection link. As the ARIS deployment affects the network topology and the large-scale channel fading, we can rewrite (\ref{eq29}) as an explicit function of the distance between the ARIS and the ground users. By introducing variables
\begin{subequations}
\begin{align}
&\boldsymbol{E}_{i_k}=[h_{i_k}^{\dagger};\sum\limits_{j=1}^J\chi_{j,k}\boldsymbol{\hat{g}}_{j,i_k}^{\dagger}\boldsymbol{\Theta}_{j} \boldsymbol{G}_{j}],\\ &\boldsymbol{F}_{i_k,l}=\boldsymbol{E}_{i_k}\boldsymbol{w}_l\boldsymbol{w}_l^{\dagger}\boldsymbol{E}_{i_k}^{\dagger},\\ &\boldsymbol{D}_{j,i_k}=[1,d_{j,i_k}^{-\frac{\beta}{2}}],
\end{align}
\end{subequations}
with $\boldsymbol{\hat{g}}_{j,i_k}=\sqrt{L_0}(\sqrt{\frac\rho{\rho+1}}\boldsymbol{g}_{j,i_k}^\mathrm{Los}+\sqrt{\frac1{\rho+1}}\boldsymbol{g}_{j,i_k}^\mathrm{NLoS})$, the rate of ground user $i_k$ in (\ref{eq29}) can be rewritten as
\begin{align}\label{3.12}
R_{i_k}=&\log\left(1+\frac{\boldsymbol{D}_{j,i_k}\boldsymbol{F}_{i_k,k}\boldsymbol{D}_{j,i_k}^{\mathrm{T}}+\sigma_{i_k}^{2}}{\sum\limits_{l=1,l\neq k}^K\boldsymbol{D}_{j,i_k}\boldsymbol{F}_{i_k,l}\boldsymbol{D}_{j,i_k}^{\mathrm{T}}+\sigma_{i_k}^{2}}\right),\nonumber\\
=&\log\left(\sum\limits_{l=1}^{K}\boldsymbol{D}_{j,i_k}\boldsymbol{F}_{i_k,l}\boldsymbol{D}_{j,i_k}^{\mathrm{T}}+\sigma_{i_k}^{2}\right)\\
&-\log\left(\sum\limits_{l=1,l\neq k}^K\boldsymbol{D}_{j,i_k}\boldsymbol{F}_{i_k,l}\boldsymbol{D}_{j,i_k}^{\mathrm{T}}+\sigma_{i_k}^{2}\right)\nonumber\\
&\forall i_k\in\mathcal{I}_k,\forall k\in\mathcal{K}.\nonumber
\end{align}
Additionally, due to the large distance between the groups, the effect of ARIS on unassociated groups will be ignored. Assuming the $j$-th ARIS is associated with the $k$-th group, the problem in (\ref{p4}) can be further decomposed into the problem of the $j$-th ARIS which can be written as
\begin{subequations}\label{eq17}
\begin{align}
\max_{\boldsymbol{q}_j,\Omega_k}\quad&\Omega_k\\
\mathrm{s.t.}\quad&R_{m_k}\geq \Omega_k, \forall m_k\in \mathcal{M}_k,\label{j3}\\
&R_{e_k}\leq\Upsilon_k, \forall e_k\in\mathcal{E}_k,\label{j4}\\
&\boldsymbol{\boldsymbol{q}}_{j}\in\mathcal{X}\times\mathcal{Y}.
\end{align}
\end{subequations}

Since the rate of the ground user $i_k$ is expressed by equation (\ref{3.12}), the constraints in (\ref{j3}) and (\ref{j4}) are non-convex. This non-convexity arises from three factors, the nonlinear distance coupling between the ARIS deployment and user positions, the concave logarithmic composition involving quadratic terms, and the concave-difference structure resulting from the subtraction of logarithmic functions. To address the nonlinear distance coupling and the concave logarithmic composition, we introduce a slack variable set $\Delta_2=\{\bar{u}_{i_k}, \hat{u}_{i_k}, \bar{S}_{i_k}, \hat{S}_{i_k}\}$. Then, (\ref{j3}) and (\ref{j4}) are transformed into
\begin{subequations}\label{all16}
\begin{align}
&\bar{u}_{m_k}\leq d_{m_k}^{-\frac{\beta}{2}},\label{w1}\\
&\hat{u}_{m_k}\geq d_{m_k}^{-\frac{\beta}{2}},\label{eq24}\\
&\bar{S}_{m_k}\leq\sum\limits_{l=1}^{K}\boldsymbol{D}_{j,m_k}\boldsymbol{F}_{m_k,l}\boldsymbol{D}_{j,m_k}^{\mathrm{T}}+\sigma_{m_k}^2,\label{eq22}\\
&\hat{S}_{m_k}\geq\sum\limits_{l=1,l\neq k}^{K}\boldsymbol{D}_{j,m_k}\boldsymbol{F}_{m_k,l}\boldsymbol{D}_{j,m_k}^{\mathrm{T}}+\sigma_{m_k}^2,\label{o1}\\
& \log(\bar{S}_{m_k})-\log(\hat{S}_{m_k})\geq \Omega_k,\quad\forall m_k\in\mathcal{M}_k\label{all14}
\end{align}
\end{subequations}
and
\begin{subequations}\label{all17}
\begin{align}
&\bar{u}_{e_k}\geq d_{e_k}^{-\frac{\beta}{2}},\label{w2}\\
&\hat{u}_{e_k}\leq d_{e_k}^{-\frac{\beta}{2}},\label{eq25}\\
&\bar{S}_{e_k}\geq\sum\limits_{l=1}^K\boldsymbol{D}_{j,e_k}\boldsymbol{F}_{e_k,l}\boldsymbol{D}_{j,e_k}^{\mathrm{T}}+\sigma_{e_k}^2,\label{o2}\\
&\hat{S}_{e_k}\leq\sum\limits_{l=1,l\neq k}^K\boldsymbol{D}_{j,e_k}\boldsymbol{F}_{e_k,l}\boldsymbol{\boldsymbol{D}}_{j,e_k}^{\mathrm{T}}+\sigma_{e_k}^2,\label{eq23}\\
 &\log(\bar{S}_{e_k})-\log(\hat{S}_{e_k})\leq \Upsilon_k,\quad\forall e_k\in\mathcal{E}_k.\label{all15}
\end{align}
\end{subequations}
Meanwhile, we can further expand the constraints (\ref{w1}), (\ref{eq24}), (\ref{w2}) and (\ref{eq25}) as
\begin{subequations}\small\label{eq26}
\begin{align}
\begin{split}
\bar{u}_{m_k}^{-\frac4{\beta}}-x_j^2-x_{m_k}^2-y_j^2-y_{m_k}^2-H^2+2(x_{m_k}x_j+y_{m_k}y_j)\geq0,
\end{split}\\
\begin{split}
  x_j^2+x_{m_k}^2+y_j^2+y_{m_k}^2+H^2-2(x_{m_k}x_j+y_{m_k}y_j)-\hat{u}^{-\frac{4}{\beta}}\geq0,
 \end{split}\\
 \begin{split}
 x_j^2+x_{m_k}^2+y_j^2+y_{m_k}^2+H^2-2(x_{m_k}x_j+y_{m_k}y_j)-\bar{u}_{e,k}^{-\frac{4}{\beta}}\geq0,
\end{split}\\
\begin{split}
\hat{u}_{e_k}^{-\frac4{\beta}}-x_j^2-x_{e_k}^2-y_j^2-y_{e_k}^2-H^2+2(x_{e_k}x_j+y_{e_k}y_j)\geq0,
\end{split}
 \end{align}
\end{subequations}
respectively. We can readily see that the constraints (\ref{eq22}), (\ref{all14}), (\ref{eq23}), (\ref{all15}) and (\ref{eq26}) are non-convex. To address this issue, the SCA method is employed. By replacing these non-convex terms with their FTSs, these constraints are rewritten as 
\begin{subequations}\label{o3}
\begin{align}
&\bar{S}_{m_k} \leq \sum_{l=1}^k \left( \boldsymbol{D}_{j,m_k} \boldsymbol{F}_{m_k,l} \boldsymbol{D}_{j,m_k}^\mathrm{T} \right)^\mathrm{lb} + \sigma_{m_k}^2, \\
&\log(\bar{S}_{m_k}) - ( \log(\hat{S}_{m_k}) )^\mathrm{ub} \geq \Omega_k, \\
&\hat{S}_{e_k} \leq \sum_{l=1, l \neq k}^K \left( \boldsymbol{D}_{j,e_k} \boldsymbol{F}_{e_k,l} \boldsymbol{D}_{j,e_k}^\mathrm{T} \right)^\mathrm{lb} + \sigma_{e_k}^2, \\
&( \log(\bar{S}_{e_k}))^\mathrm{ub} - \log(\hat{S}_{e_k}) \leq \Upsilon_k, \\
&(\bar{u}_{m_k}^{-\frac{4}{\beta}})^\mathrm{lb} - x_j^2 - x_{m_k}^2 - y_j^2 - y_{m_k}^2 \nonumber \\
& \hspace{2cm}- H^2 + 2(x_{m_k} x_j + y_{m_k} y_j) \geq 0, \\
&( x_j^2 )^\mathrm{lb} + \left( x_{m_k}^2 \right)^\mathrm{lb} + y_j^2 + y_{m_k}^2 + H^2 \nonumber \\
& \hspace{2cm}- 2(x_{m_k} x_j + y_{m_k} y_j) - \hat{u}^{-\frac{4}{\beta}} \geq 0, \\
&( x_j^2 )^\mathrm{lb} + x_{m_k}^2 + ( y_j^2 )^\mathrm{lb} + y_{m_k}^2 + H^2 \nonumber \\
&\hspace{2cm}- 2(x_{m_k} x_j + y_{m_k} y_j) - \bar{u}_{e,k}^{-\frac{4}{\beta}} \geq 0, \\
&( \hat{u}_{e_k}^{-\frac{4}{\beta}})^\mathrm{lb} - x_j^2- x_{e_k}^2 - y_j^2 - y_{e_k}^2 \nonumber \\
& \hspace{2cm}- H^2 + 2(x_{e_k} x_j + y_{e_k} y_j) \geq 0,
\end{align}
\end{subequations}
respectively. The FTSs of these terms are given in the Appendix. Finally, problem (\ref{eq17}) is recast as the following convex one:
\begin{subequations}\label{all19}
 \begin{align}
\max_{\boldsymbol{q}_{j},\Omega_k,\Delta_2}\quad &\Omega_k\\
\mathrm{s.t.}\quad&\boldsymbol{\boldsymbol{q}}_j\in\mathcal{X}\times\mathcal{Y},\quad\forall j\in\mathcal{J},\\
&(\ref{o1}), (\ref{o2}), (\ref{o3}),\nonumber
\end{align}
\end{subequations}
where the constraints in (\ref{o1}), (\ref{o2}), and (\ref{o3}) are the convexified counterparts of (\ref{j3}) and (\ref{j4}). The problem in (\ref{all19}), as an approximated version of (\ref{p3}), incorporates a linear objective function and all convex constraints. Then, the original deployment optimization can be tackled by using the SCA method.

\subsection{Overall Algorithm Design}
\begin{algorithm}[!ht]\small
    \caption{Multi-ARIS Assisted Secure MSS}
    \renewcommand{\algorithmicrequire}{\textbf{Input:}}
    \renewcommand{\algorithmicensure}{\textbf{Output:}}
    \begin{algorithmic}[1]
        \Require Network topology, channel parameters, stopping thresholds $\epsilon_S, \epsilon_B, \epsilon_G$.
        \State Initialize $t\leftarrow 0$, and feasible points $\boldsymbol{w}_k^{(0)}, \boldsymbol{\theta}_j^{(0)}, \chi_{j,k}^{(0)}, \boldsymbol{q}_j^{(0)}$;
        \Repeat
            \State $t\leftarrow t+1$;
            
            \State $n \leftarrow 0$; $\boldsymbol{w}_k^{(n=0)} \leftarrow \boldsymbol{w}_k^{(t-1)}$; $\boldsymbol{\theta}_j^{(n=0)} \leftarrow \boldsymbol{\theta}_j^{(t-1)}$;
            \Repeat
                \State $n \leftarrow n+1$;
                \State \Comment{Update transmitter beamforming $\boldsymbol{w}_k$}
                \State $\ell\leftarrow 0$; $\boldsymbol{w}_k^{(\ell)} \leftarrow \boldsymbol{w}_k^{(n-1)}$;
                \Repeat
                    \State $\ell\leftarrow \ell+1$; Calculate SCA parameters $\xi_{m_k}^{(\ell)}, \xi_{e_k}^{(\ell)}$;
                    \State Update $\boldsymbol{w}_k^{(\ell)}$ using (\ref{eq13});
                \Until $\|\boldsymbol{w}_k^{(\ell)}-\boldsymbol{w}_k^{(\ell-1)}\|\leq\epsilon_T$;
                \State $\boldsymbol{w}_k^{(n)} \leftarrow \boldsymbol{w}_k^{(\ell)}$;

                \State \Comment{Update ARIS phase shifts $\boldsymbol{\theta}_j$}
                \For{$j=1,\dots,J$}
                    \State $\ell\leftarrow 0$; $\boldsymbol{\theta}_j^{(\ell)} \leftarrow \boldsymbol{\theta}_j^{(n-1)}$;
                    \Repeat
                        \State $\ell\leftarrow \ell+1$; 
                        \State Calculate SCA parameters $\mu_{m_k}^{(\ell)}, \mu_{e_k}^{(\ell)}$;
                        \State Update $\boldsymbol{\theta}_j^{(\ell)}$ using (\ref{eq16});
                    \Until $\|\boldsymbol{\theta}_j^{(\ell)}-\boldsymbol{\theta}_j^{(\ell-1)}\|\leq\epsilon_R$;
                    \State $\boldsymbol{\theta}_j^{(n)} \leftarrow \boldsymbol{\theta}_j^{(\ell)}$;
                \EndFor              
            \Until$\frac{\| [\text{vec}(\boldsymbol{w}_k^{(n)}), \text{vec}(\boldsymbol{\theta}_j^{(n)})] - [\text{vec}(\boldsymbol{w}_k^{(n-1)}), \text{vec}(\boldsymbol{\theta}_j^{(n-1)})] \|}{\| [\text{vec}(\boldsymbol{w}_k^{(n-1)}), \text{vec}(\boldsymbol{\theta}_j^{(n-1)})] \|} \leq \epsilon_S$;
            \State Set $\boldsymbol{w}_k^{(t)} \leftarrow \boldsymbol{w}_k^{(n)}$; $\boldsymbol{\theta}_j^{(t)} \leftarrow \boldsymbol{\theta}_j^{(n)}$;

            \State $n \leftarrow 0$; $\chi_{j,k}^{(n=0)} \leftarrow \chi_{j,k}^{(t-1)}$; $\boldsymbol{q}_j^{(n=0)} \leftarrow \boldsymbol{q}_j^{(t-1)}$;
            \Repeat
                \State $n \leftarrow n+1$;
                \State \Comment{Update ARIS association $\chi_{j,k}$}
                \State $\ell\leftarrow 0$; $\chi_{j,k}^{(\ell)} \leftarrow \chi_{j,k}^{(n-1)}$;
                \Repeat
                    \State $\ell\leftarrow \ell+1$; Update $\chi_{j,k}^{(\ell)}, \Delta_1^{(\ell)}$;
                \Until {$\max_{j,k}|\chi_{j,k}^{(\ell)}-\chi_{j,k}^{(\ell-1)}|\leq\epsilon_A$};
                \State $\chi_{j,k}^{(n)} \leftarrow \chi_{j,k}^{(\ell)}$;

                \State \Comment{Update ARIS deployment $\boldsymbol{q}_j$}
                \For{$j=1,\dots,J$}
                    \State $\ell\leftarrow 0$; $\boldsymbol{q}_j^{(\ell)} \leftarrow \boldsymbol{q}_j^{(n-1)}$;
                    \Repeat
                        \State $\ell\leftarrow \ell+1$; Update $\boldsymbol{q}_{j}^{(\ell)}, \Delta_2^{(\ell)}$ based on $\chi_{j,k}^{(n)}$;
                    \Until $\|\boldsymbol{q}_j^{(\ell)}-\boldsymbol{q}_j^{(\ell-1)}\|\leq\epsilon_D$;
                    \State $\boldsymbol{q}_j^{(n)} \leftarrow \boldsymbol{q}_j^{(\ell)}$;
                \EndFor
            \Until$\frac{\| [\text{vec}(\chi_{j,k}^{(n)}), \text{vec}(\boldsymbol{q}_j^{(n)})] - [\text{vec}(\chi_{j,k}^{(n-1)}), \text{vec}(\boldsymbol{q}_j^{(n-1)})] \|}{\| [\text{vec}(\chi_{j,k}^{(n-1)}), \text{vec}(\boldsymbol{q}_j^{(n-1)})] \|} \leq \epsilon_L$;
            \State Set $\chi_{j,k}^{(t)} \leftarrow \chi_{j,k}^{(n)}$; $\boldsymbol{q}_j^{(t)} \leftarrow \boldsymbol{q}_j^{(n)}$;

        \Until Convergence
        \Ensure Optimized variables $\boldsymbol{w}_k^{(t)}, \boldsymbol{\theta}_j^{(t)}, \chi_{j,k}^{(t)}, \boldsymbol{q}_j^{(t)}$.
    \end{algorithmic}
\end{algorithm}

Based on the above discussion, we first address the non-continuity issue arising from the minimization of the objective function by introducing auxiliary variables. Then, we decompose the problem into small-scale and large-scale subproblems. The solution to problem (\ref{eq5}) can be obtained by solving the large-scale and small-scale subproblems within a BCD framework. Therefore, the algorithm toward Multi-ARIS Assisted Secure MSS is summarized in Algorithm 1, where $t=0,1,2,\dots,$ denotes the iterations and the constant $\epsilon$ determines the overall convergence. The convergence threshold for the small-scale problem is $\epsilon_S$, while the threshold for the large-scale problem is $\epsilon_B$. The transmission beamforming and passive beamforming require inner loops in the form of alternating optimization with auxiliary variables, where the constants $\epsilon_T$ and $\epsilon_R$ indicate the convergence. The ARIS association and deployment require inner loops in the form of SCA procedures, where the constants $\epsilon_A$ and $\epsilon_D$ indicate the convergence. Since the maximum transmit power is limited, the proposed algorithm is upper bounded and guaranteed to converge.

In Algorithm 1, lines 5-25 correspond to the small-scale subproblem, which includes transmission beamforming and passive beamforming. Specifically, lines 9-14 address transmission beamforming, and lines 16-23 focus on passive beamforming. Meanwhile, lines 27-44 correspond to the large-scale subproblem, which covers ARIS association and deployment. Lines 31-35 deal with ARIS association, and lines 37-42 address ARIS deployment. 

Moreover, we briefly analyze the computation complexity of the proposed algorithm. For the transmission beamforming, the semidefinite programming is of a complexity of $\mathcal{O}\bigl( (KL^2)^{4.5}\bigr)$ and we assume there are $I_1$ times of iterations, then the incurred complexity is $\mathcal{O}\bigl(I_1(KL^2)^{4.5}\bigr)$. Similarly, solving for the passive beamforming requires a complexity of $\mathcal{O}\Bigl(J I_2 \bigl((N+1)^2\bigr)^{4.5}\Bigr)$, where $I_2$ is the number of iterations. For the ARIS association, the SCA method is of a complexity of  $\mathcal{O}\biggl(\Bigl(JK + \sum\limits_{k=1}^K (M_k + E_k)\Bigr)^{3.5}\biggr)$ and we assume there are $I_3$ times of iterations, then the incurred complexity is $\mathcal{O}\biggl(I_3 \Bigl(JK + \sum\limits_{k=1}^K (M_k + E_k)\Bigr)^{3.5}\biggr)$. Similarly, solving for the ARIS deployment requires a complexity of $\mathcal{O}\biggl(J I_4\Bigl(\max\limits_{k\in\mathcal{K}}(M_k + E_k)\Bigr)^{3.5}\biggr)$, where $I_4$ is the number of iterations. Finally, we assume the iterations of  the small-scale optimization problem are of $I_S$ times, the iterations of  the large-scale optimization problem are of $I_L$ times and the outer iterations among the small-scale optimization problem and the large-scale optimization problem are of $I_0$ times, then the overall complexity is $\mathcal{O}\Biggl(I_0\Biggl[I_S\Bigl(I_1 \cdot (KL^2)^{4.5} + J I_2\bigl((N+1)^2\bigr)^{4.5}\Bigr) + I_L\biggl(I_3 \Bigl(JK + \sum\limits_{k=1}^K (M_k + E_k)\Bigr)^{3.5} + J  I_4\bigl(\max\limits_{k\in\mathcal{K}} (M_k + E_k)\bigr)^{3.5}\biggr)\Biggr]\Biggr)$.

\section{Simulation Results}\label{s6}
In this section, numerical results are provided to demonstrate the performance of the proposed multi-ARIS assisted satellite systems. We consider the downlink transmission from a multibeam satellite in low-earth-orbit (LEO) at an altitude of 220 km to ground users. The communication takes a carrier frequency of 6 GHz, and the Doppler effect caused by satellite movements is assumed to be compensated. The processing bandwidth and noise temperature are set to $B= 200$ MHz and $T_{i_k}= 290$ K, $\forall i_k\in\mathcal{I}_k, \forall k\in\mathcal{K}$, respectively. The number of ARISs is $J=3$, and the number of groups is $K=5$. Meanwhile, we assume that there are three legitimate users and one eavesdropper within each group. The number of satellite antennas is $L=5$, the total satellite power is $P_T= 100$ W. The maximum antenna gain of the satellite is 50 dBi. The feed radiation pattern stems from the European Space Agency (ESA) as adopted in~\cite{a2}. The radius of the group is 300 m, with the ground users and eavesdroppers randomly distributed within the group. The ARISs are deployed at an altitude of 100 m. Each ARIS consists of 25 subsurfaces, and there are 25 elements by default in each subsurface to cope with long-distance path loss. Each ARIS element separation and wavelength induce a ratio of 0.5. The path loss at the reference distance of 1 m is 20 dB. The air-ground channels experience a path loss exponent of 2.3. The Rician factor is set to 3 dB. The penalty factor $\tau=10$. The scenario configurations used in~\cite{a12,a27,a39} are referenced to establish the simulation settings in this work.

We set $\epsilon_T = \epsilon_R = 1 \times 10^{-3}$ as the convergence threshold for the inner loops of transmission beamforming and passive beamforming subproblems. Similarly, $\epsilon_A = \epsilon_D = 1 \times 10^{-2}$ is set as the convergence threshold for the inner loops of  ARIS association and ARIS deployment subproblems. Additionally, $\epsilon_S  = 1 \times 10^{-3}$ and $\epsilon_L=1 \times 10^{-2}$ are set as the convergence thresholds for the small-scale and large-scale subproblems, respectively. Meanwhile, $ \epsilon = 1 \times 10^{-3}$ is used as the outer loop for the overall problem, respectively. In each independent simulation, we randomly generate the positions of users and eavesdroppers within each group. The presented simulation performance represents the average results of numerous simulations.

\begin{figure}[h!]
    \centering
    \includegraphics[trim=100 270 100 270, clip, width=0.72\linewidth]{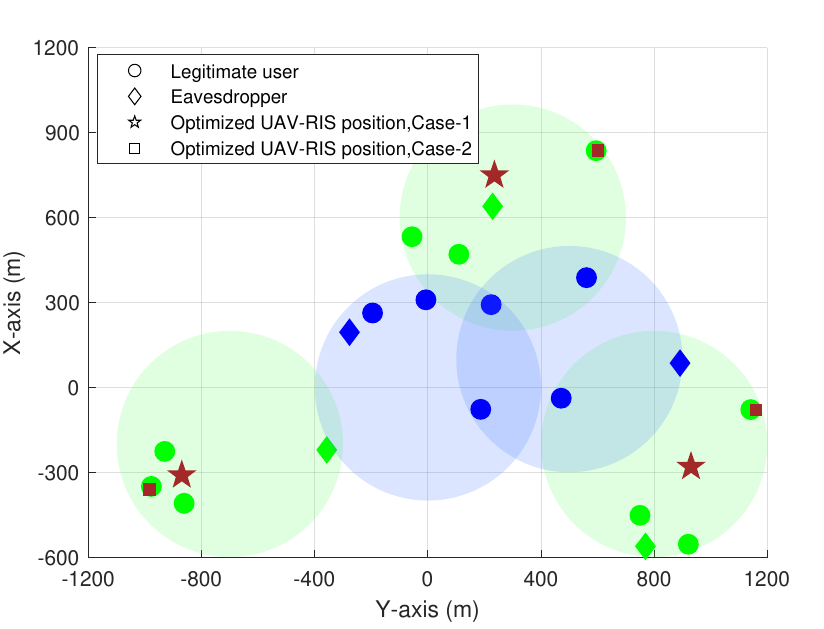}
    \caption{ARIS association and deployment in the presence of overlapping user groups}
    \label{fig1}
\end{figure}
\begin{figure}[h!]
    \centering
    \includegraphics[trim=100 265 100 265, clip, width=0.72\linewidth]{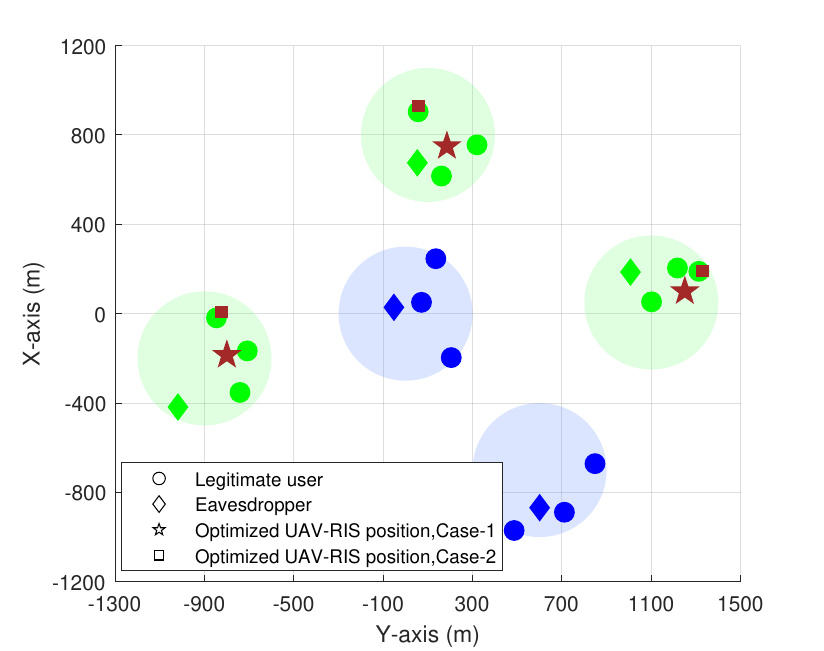}
    \caption{ARIS association and deployment in the presence of independent user groups}
    \label{fig2}
\end{figure}
Fig. \ref{fig1} and Fig. \ref{fig2} depict the ARIS association and the ARIS deployment. The groups with green color represent those that are associated with the ARIS, while the groups with blue color indicate those that have not been associated with the ARIS. It is evident that the ARIS system is predominantly associated with groups located at the edges of all groups. This phenomenon occurs because groups located closer to the network center experience stronger inter-beam interference, resulting in reduced rates for both legitimate users and eavesdroppers. Consequently, ARISs are preferentially deployed at the edge of the network, where the inter-beam interference is lower, enhancing both system performance and security. Meanwhile, in Case 1, the ARISs are equipped with subsurfaces, and each subsurface contains 25 elements. As the quality of reflection links is significantly enhanced by numerous elements in the subsurface, the ARIS is often deployed near the geometric center of the users within the group. Then the users within the same group are able to achieve the same achievable rate. In Case 2, we assume that the ARIS is not equipped with subsurfaces. In this scenario, the location of the ARISs are closer to the user with the lowest achievable rate compared to other users within the group.

In the following simulations, we adopt two benchmark schemes: "without RIS" and "ARIS with Fixed Deployment". For the benchmark of "without RIS", there are no ARIS in the system and communication relies on the direct link. To ensure a fair comparison, this benchmark scheme still utilizes the beamforming optimization proposed in our algorithm. For the benchmark of "ARIS Fixed Deployment" scheme, the deployment locations of the ARIS are fixed, while the beamforming and ARIS association are still optimized to ensure a fair comparison.

\begin{figure}[h!]
    \centering
    \includegraphics[trim= 95 270 95 270, clip, width=0.72\linewidth]{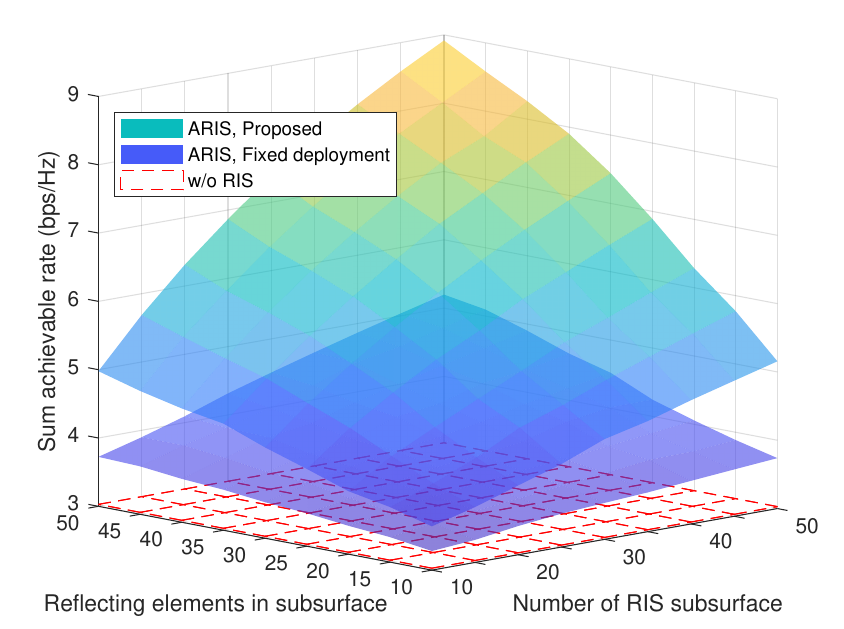}
    \caption{Achievable sum rate versus the number of reflecting elements in each subsurface and the number of ARIS subsurfaces.}
    \label{fig3}
\end{figure}

In Fig. \ref{fig3}, we dramatically increase the number of reflecting elements  in each subsurface and the number of ARIS subsurfaces to their influence on the performance of the considered system. It can be seen that  the number of reflecting elements  in each subsurface is an essential factor affecting the performance of the system. As the number of reflecting elements  in each subsurface  increases, the performance is significantly improved by fully exploiting the increased number of elements in each ARIS. Therefore, incorporating subsurfaces into ARIS is crucial for optimizing performance. Meanwhile, the observed performance gap compared to fixed RIS deployment demonstrates the necessity of optimizing ARIS deployment. The static positioning of fixed RIS constrains their channel reconfiguration capability. In contrast, the proposed scheme optimizes ARIS deployment based on the locations of the ground nodes. This adaptability alters the network topology, enhancing channel conditions, while coordinated beamforming improves system performance. Therefore, optimizing ARIS placement is critical for multi-ARIS systems.

\begin{figure}[h!]
    \centering
    \includegraphics[trim=85 260 85 260, clip, width=0.72\linewidth]{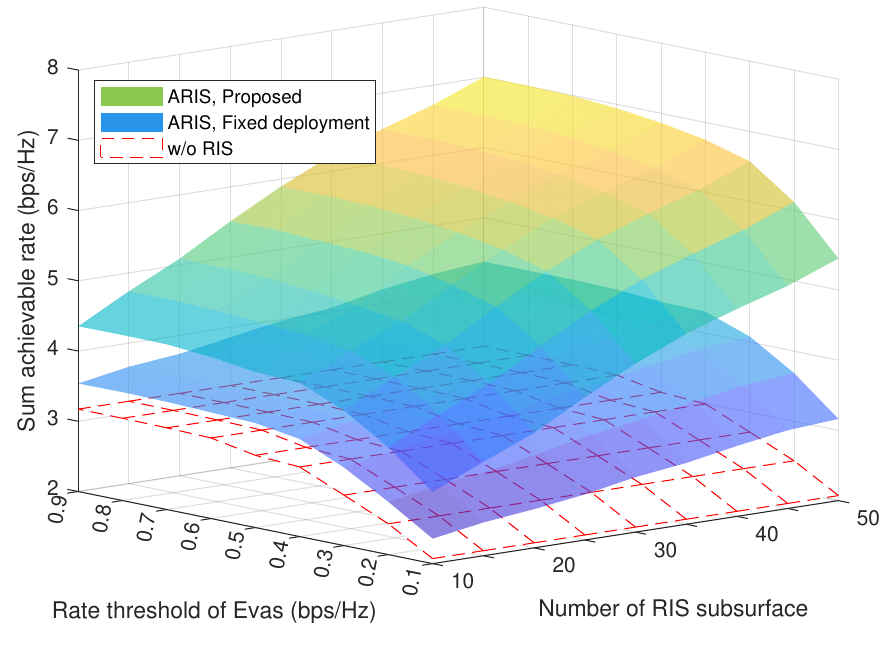}
    \caption{Achievable sum rate versus rate threshold of Eves and number of ARIS subsurface. }
    \label{fig4}
\end{figure}

In addition, to consider the influence of the eavesdroppers' constraints  on the system performance, we simulated the variation of the sum rate, as shown in Fig. \ref{fig4}. The achievable sum rate increases with the growing eavesdroppers' rate threshold and the increasing number of ARIS subsurfaces. This is because when the constraint of the eavesdropper rate threshold is relaxed, the optimization algorithm gains an expanded feasible region, allowing it to more effectively maximize the communication rate for legitimate users while still satisfying the security requirements. To provide further insights, the proposed scheme’s benefits over the fixed RISs by optimizing the ARIS association and the ARISs deployment. In the proposed ARIS-assisted scheme, the direct and reflected paths are combined constructively at the legitimate user while combined destructively at the eavesdroppers, which allows us to increase the achievable sum rate while limiting the eavesdroppers' rate. Meanwhile, it has been observed that when $\Upsilon_k$ exceeds 0.7 bps/Hz, the achievable sum rate becomes independent of further increases in  $\Upsilon_k$. This trend can be attributed to the fact that 0.7 bps/Hz corresponds to the maximum interceptable rate for the most capable eavesdropper at that particular location. When this maximum interceptable rate is exceeded, the system's performance is no longer constrained by the eavesdropper's threshold but is limited by the total transmit power and the channel conditions. Therefore, even if the security threshold is further relaxed, the sum rate cannot be further improved.

\begin{figure}[h!]
    \centering
    \includegraphics[trim=100 270 100 270, clip, width=0.72\linewidth]{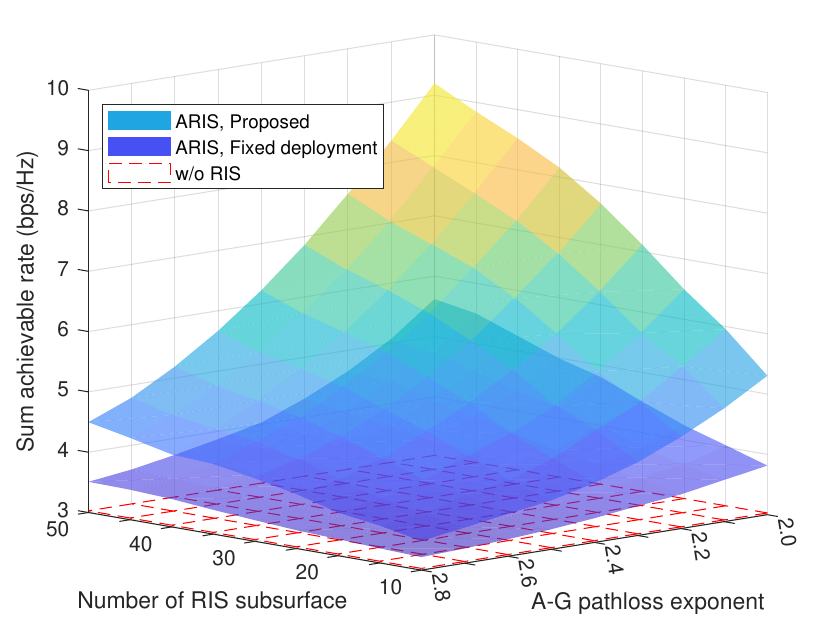}
    \caption{Achievable sum rate versus path loss exponent and number of ARIS subsurface. }
    \label{fig5}
\end{figure}
\begin{figure}[h!]
    \centering
    \includegraphics[trim=100 270 100 270, clip, width=0.72\linewidth]{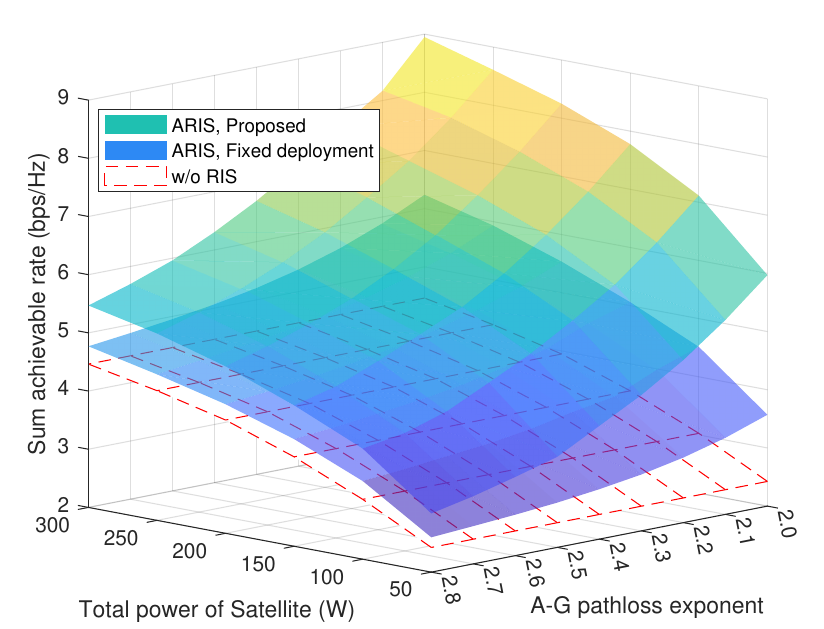}
    \caption{Achievable sum rate versus satellite transmission power and number of ARIS subsurface. }
    \label{fig6}
\end{figure}
In Fig. \ref{fig5}, we present the achievable sum rate versus the path loss exponent and the number of ARIS subsurfaces. It has been observed that the path loss exponent of reflection links substantially impacts system performance. Specifically, as the path loss exponent decreases, the achievable sum rate increases. This is because a lower path loss exponent leads to higher-quality reflection links, subsequently enhancing the achievable sum rate. It is observed that with the increasing number of reflecting elements, the sum rate exhibits an upward trend. This is because the increased number of reflecting elements can effectively compensate for high large-scale path loss. However, it is observed that increasing the number of ARIS subsurface leads to marginally diminishing returns in system performance gain. This trend occurs because increasing the number of subsurfaces simultaneously enhances the legitimate link and the eavesdropper's channel, thereby imposing a theoretical upper bound on physical-layer security performance. Meanwhile, Fig. \ref{fig6} shows the achievable sum rate versus the path loss exponent and the satellite transmission power. The impact of the path loss exponent of the reflection links on the system is similar to that in Fig. \ref{fig5}. Furthermore, by comparing Fig. \ref{fig5} and Fig. \ref{fig6}, it can be observed that increasing transmission power provides better compensation for the reflection link than increasing the number of ARIS reflection elements. This is because increasing the transmission power not only improves the quality of the reflection link, but also enhances the quality of the direct link. This phenomenon indicates that increasing satellite transmit power provides a systemic enhancement, strengthening not only the reflected link but also the direct link. In contrast, although increasing the number of ARIS reflection elements can enhance the quality of reflection link, the effectiveness is limited by the characteristics of the reflection link, and it cannot provide the same performance of compensation as an increase in transmission power.

\begin{figure}[h!]
    \centering
    \includegraphics[trim=100 270 100 270, clip, width=0.72\linewidth]{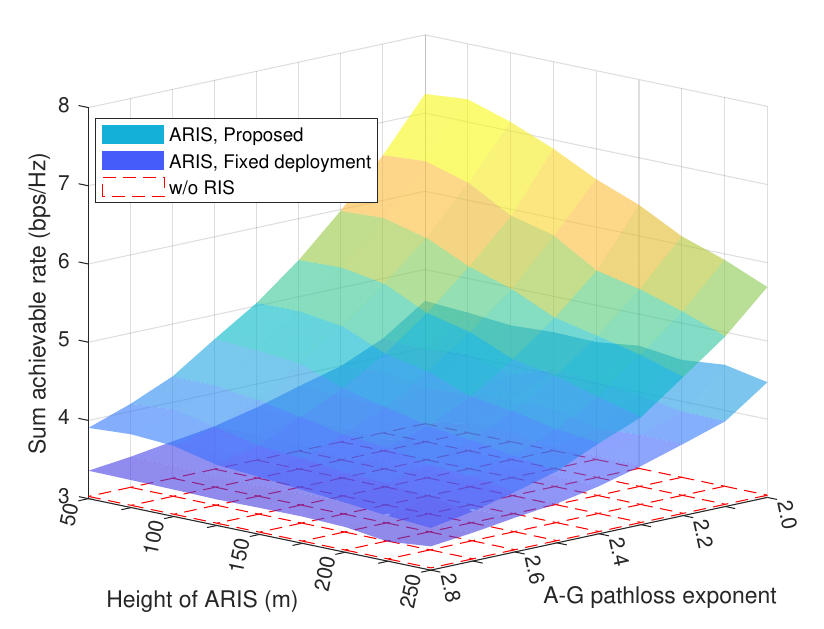}
    \caption{Achievable sum rate versus altitude of ARIS and  path loss exponent.}
    \label{fig7}
\end{figure}
Fig. \ref{fig7} illustrates the achievable sum rate versus the altitude of the ARIS and the path-loss exponent. The influence of the path-loss exponent on system performance corresponds to that depicted in Fig.\ref{fig5}. Meanwhile, the performance of the fixed RIS is less dependent on its height. This is because that the distance between the fixed RIS and the users ranges from 500 meters to 700 meters, the variations in the altitude of the fixed RIS have a minimal impact on the quality of the reflection link. In contrast, the performance of our proposed optimization scheme is significantly sensitive to the ARIS altitude. This is because the ARIS can be deployed at the optimal horizontal position, where the ARIS altitude becomes a key factor influencing the reflection link quality. When the path loss exponent is fixed, a lower ARIS deployment altitude can effectively enhance link quality and improve the sum achievable rate.

\section{Conclusion}\label{s7}
In this paper, we investigate the physical layer security of ARIS-assisted multibeam satellite systems. To solve this problem, we propose an algorithm based on the BCD framework to maximize the achievable sum rate. The transmission beamforming and passive beamforming are optimized using semidefinite programming. The ARIS association and ARIS deployment are solved by successive convex approximation. The simulation results demonstrate that strategically using ARIS can markedly improve the achievable sum rate, especially considering the constraints of eavesdroppers' rate. This finding provides significant insights and innovative methodologies for enhancing security technologies in future satellite communications.

\section*{Appendix}
\subsection{The Approximation in Section \ref{ap1}}
For the given point $\{\chi_{j,k}^{(\ell)},\boldsymbol{\chi}_k^{(\ell)},\hat{\psi}_{m_k}^{(\ell)},\bar{\psi}_{e_k}^{(\ell)}\}$, the FTSs of  $\chi_{j,k}-(\chi_{j,k})^2$, $\boldsymbol{\chi}_k\boldsymbol{C}_{m_k,l}\boldsymbol{\chi}_k^{\mathrm{T}}$, $\boldsymbol{\chi}_k\boldsymbol{C}_{e_k,l}\boldsymbol{\chi}_k^{\mathrm{T}}$, $\log(\hat{\psi}_{m_k})$, $\log(\bar{\psi}_{e_k})$ are given as
\begin{subequations}\label{all11}
\begin{align}
(\chi_{j,k}-(\chi_{j,k})^2)^{\mathrm{ub}}=\chi_{j,k}-(\chi_{j,k}^{\mathrm{(\ell)}})^2-2\chi_{j,k}^{(\ell)}(\chi_{j,k}-\chi_{j,k}^{(\ell)}).
\end{align}
\vspace{-\baselineskip} 
    \begin{align}
(\boldsymbol{\chi}_k\boldsymbol{C}_{m_k,l}\boldsymbol{\chi}_k^{\mathrm{T}})^\mathrm{lb} =& -\boldsymbol{\chi}_{k}^{(\ell)}\boldsymbol{C}_{m_k,l}(\boldsymbol{\chi}_{k}^{(\ell)})^{\mathrm{T}} \nonumber \\
    \quad &+2 \Re\left\{\boldsymbol{\chi}_{k}^{(\ell)} \boldsymbol{C}_{m_k,l} (\boldsymbol{\chi}_{k})^{\mathrm{T}}\right\}, \\
(\boldsymbol{\chi}_{k}\boldsymbol{C}_{e_k,l}\boldsymbol{\chi}_{k}^\mathrm{T})^\mathrm{lb} =& -\boldsymbol{\chi}_{k}^{(\ell)}\boldsymbol{C}_{e_k,l} (\boldsymbol{\chi}_{k}^{(\ell)})^{\mathrm{T}} \nonumber \\
    \quad &+ 2 \Re\left\{\boldsymbol{\chi}_{k}^{(\ell)} \boldsymbol{C}_{e_k,l} (\boldsymbol{\chi}_{k})^{\mathrm{T}}\right\}, 
    \end{align}
\vspace{-\baselineskip} 
\begin{align}
 &(\log(\hat{\psi}_{m_k}))^\mathrm{ub} = \log(\hat{\psi}_{m_k}^{(\ell)}) + (\hat{\psi}_{m_k} - \hat{\psi}_{m_k}^{(\ell)})\left(\frac{1}{\ln 2} \bar{\psi}_{m_k}^{(\ell)}\right),\\
   &(\log(\bar{\psi}_{e_k}))^\mathrm{ub} = \log(\bar{\psi}_{e_k}^{(\ell)}) + (\bar{\psi}_{e_k} - \bar{\psi}_{e_k}^{(\ell)})\left(\frac{1}{\ln 2} \bar{\psi}_{e_k}^{(\ell)}\right),
\end{align}
\end{subequations}
respectively.

\subsection{The Approximation in Section \ref{ap2}}
For the given point $\{\hat{S}_{m_k}^{(\ell)},\bar{S}_{e_k}^{(\ell)},\boldsymbol{D}_{j,m_k}^{(\ell)},\boldsymbol{D}_{j,e_k}^{(\ell)},\bar{u}_{m_k}^{(\ell)},\hat{u}_{e_k}^{(\ell)}, \boldsymbol{q}_j^{(\ell)}\}$,  the FTSs of $\log(\hat{S}_{m_k})$, $\log(\bar{S}_{e_k})$, $\boldsymbol{D}_{j,m_k}\boldsymbol{F}_{m_k,l}\boldsymbol{D}_{j,m_k}^{\mathrm{T}}$, $\boldsymbol{D}_{j,e_k}\boldsymbol{F}_{e_k,l}\boldsymbol{D}_{j,e_k}^{\mathrm{T}}$, $\bar{u}_{m_k}^{-\frac{4}{\beta}}$, $\hat{u}_{e_k}^{-\frac{4}{\beta}}$, $x_j^2$ and $y_j^2$ are given as
\begin{subequations}\label{eq28}
\begin{align}
(\boldsymbol{D}_{j,m_k}\boldsymbol{F}_{m_k,l}\boldsymbol{D}_{j,m_k}^\mathrm{T})^\mathrm{lb}=&-\boldsymbol{D}_{j,m_k}^{(\ell)}\boldsymbol{F}_{m_k,l}(\boldsymbol{D}_{j,m_k}^{(\ell)})^{\mathrm{T}} \nonumber \\
    \quad &+2 \Re\left\{\boldsymbol{D}_{j,m_k}^{(\ell)} \boldsymbol{F}_{m_k,l} (\boldsymbol{D}_{j,m_k})^{\mathrm{T}}\right\}, \\
(\boldsymbol{D}_{j,e_k}\boldsymbol{F}_{e_k,l}\boldsymbol{D}_{j,e_k}^\mathrm{T})^\mathrm{lb} = &-\boldsymbol{D}_{j,e_k}^{(\ell)}\boldsymbol{F}_{e_k,l} (\boldsymbol{D}_{j,e_k}^{(\ell)})^{\mathrm{T}} \nonumber \\
    \quad &+ 2 \Re\left\{\boldsymbol{D}_{j,e_k}^{(\ell)} \boldsymbol{F}_{e_k,l} (\boldsymbol{D}_{j,e_k})^{\mathrm{T}}\right\}, 
    \end{align}
\vspace{-\baselineskip} 
\begin{align}
 &(\log(\hat{S}_{m_k}))^\mathrm{ub} = \log(\hat{S}_{m_k}^{(\ell)}) + (\hat{S}_{m_k} - \hat{S}_{m_k}^{(\ell)})\left(\frac{1}{\ln 2} \bar{S}_{m_k}^{(\ell)}\right),\\
   &(\log(\bar{S}_{e_k}))^\mathrm{ub} = \log(\bar{S}_{e_k}^{(\ell)}) + (\bar{S}_{e_k} - \bar{S}_{e_k}^{(\ell)})\left(\frac{1}{\ln 2} \bar{S}_{e_k}^{(\ell)}\right), 
\end{align}
    \vspace{-\baselineskip} 
    \begin{align}
        &(\bar{u}_{m_k}^{-\frac{4}{\beta}})^{\mathrm{lb}} = (\bar{u}_{m_k}^{(\ell)})^{-\frac{4}{\beta}}-\frac{4}{\beta}(\bar{u}_{m_k}-\bar{u}_{m_k}^{(\ell)})(\bar{u}_{m_k}^{(\ell)})^{-\frac{4}{\beta}-1},\\
        &(\hat{u}_{e_k}^{-\frac{4}{\beta}})^{\mathrm{lb}} = (\hat{u}_{e_k}^{(\ell)})^{-\frac{4}{\beta}}-\frac{4}{\beta}(\hat{u}_{e_k}-\hat{u}_{e_k}^{(\ell)})(\hat{u}_{e_k}^{(\ell)})^{-\frac{4}{\beta}-1},
    \end{align}
    \vspace{-\baselineskip} 
\begin{align}
 (x_j^2)^\mathrm{lb}& = -(x_j^{(\ell)})^2 + 2 x_j^{(\ell)} x_j, \\
(y_j^2)^\mathrm{lb}& = -(y_j^{(\ell)})^2 + 2 y_j^{(\ell)} y_j,
\end{align}  
\end{subequations}
respectively.

\vfill
\end{document}